\documentclass[journal]{IEEEtran}
\UseRawInputEncoding 
\usepackage{cite}
\usepackage{url}
\PassOptionsToPackage{hyphens}{url}\usepackage{hyperref}
\usepackage{amsmath,amssymb,amsfonts}
\usepackage{algorithmic}
\usepackage{graphicx}
\usepackage{textcomp}
\usepackage{xcolor}
\usepackage{float}
\usepackage{caption}
\usepackage{dblfloatfix}
\usepackage{booktabs}
\usepackage{multirow}
\usepackage{musicography}
\usepackage{graphicx}
\usepackage[caption=false,font=normalsize,labelfont=sf,textfont=sf]{subfig}
\usepackage{comment}

\usepackage{xcolor}

\newcommand{\iantext}[1]{{#1}}

\begin{document}

\title{Theme Transformer: Symbolic Music Generation with Theme-Conditioned Transformer}

\author{Yi-Jen Shih, Shih-Lun Wu, Frank Zalkow, Meinard M\"uller, and Yi-Hsuan Yang
}

\markboth{Journal of \LaTeX\ Class Files,~Vol.~14, No.~8, August~2021}%
{Shell \MakeLowercase{\textit{et al.}}: A Sample Article Using IEEEtran.cls for IEEE Journals}


\maketitle

\begin{abstract}
Attention-based Transformer models have been increasingly employed for automatic music generation.
To condition 
the generation process of such a model with a user-specified sequence,
a popular approach is to take that conditioning sequence as a priming sequence 
and ask a Transformer decoder to generate a continuation. 
However, 
this \emph{prompt-based conditioning} 
cannot guarantee that the conditioning sequence would develop or even simply repeat itself 
in the generated continuation. 
In this paper, 
we propose an alternative conditioning approach, called \emph{theme-based conditioning}, that explicitly trains the Transformer to treat the conditioning sequence as a thematic material that has to manifest itself multiple times 
in its generation result.  This is achieved with two main technical contributions. 
First, we propose a deep learning-based approach that uses contrastive representation learning and clustering to automatically retrieve thematic materials from music pieces in the training data.
Second, we propose a novel gated parallel attention module to be used in a sequence-to-sequence (seq2seq) 
encoder/decoder  architecture to more effectively account for a given conditioning thematic material in the generation process of the Transformer decoder.
We report on objective and subjective evaluations of variants of the proposed Theme Transformer and the conventional prompt-based baseline, showing that our best model can generate, to some extent, polyphonic pop piano music with repetition and plausible variations of a given condition. 
\end{abstract}


\section{Introduction}
Speaking of songwriting, \emph{musical ideas} are usually the first thing that strikes a composer’s mind \cite{jacob96,elowsson12icmpc}.
A musical idea can 
take the form of 
a motif, a phrase or a theme.\footnote{In Western classical music, both themes and motifs can be considered short, salient recurring patterns. \emph{Themes} are musical ideas that convey a sense of ``completeness'' and
``roundedness,'' and are essential elements used to build a composition, or a part of it~\cite{ZalkowBAM20MTDTISMIR}.  In contrast, \emph{motifs} are considered to be shorter and more basic~\cite{Drabkin01ThemeGrove}. 
}
Composers can write a piece of music by developing a given musical  idea throughout the piece via repetitions and variations.

Research on automatic symbolic music generation or style transfer has seen significant progress in recent years \cite{8007229,8918424,9178446,9302731,9376975,9540852,ji20survey}.  
Among such efforts, the adoption of the Transformer decoder-based neural network  \cite{vaswani2017attention}  
as the backbone generative model has become popular \cite{huang2018music,payne2019musenet,donahue2019lakhnes,9132664,huang20remitransformer,ren2020popmag,wu20arxivTransformerXL,ens20arxiv,wang20ismir,hsiao21aaai,muhamed21aaai,songmass}. 
The fame of the Transformer decoder can be attributed to its ``self-attention'' mechanism, which provides the generative model a much longer \emph{memory} compared to  previous statistical 
or recurrent neural network (RNN) models. 
A Transformer decoder can 
take a random seed 
and generate a  piece from scratch, or take any user-provided music fragment as the ``prompt'' and generate a continuation \cite{huang2018music}.
In either case, the long-term memory contributes to the generated piece's overall harmonic and stylistic coherence.

\begin{figure}[t]
\centering
\includegraphics[width=\columnwidth]{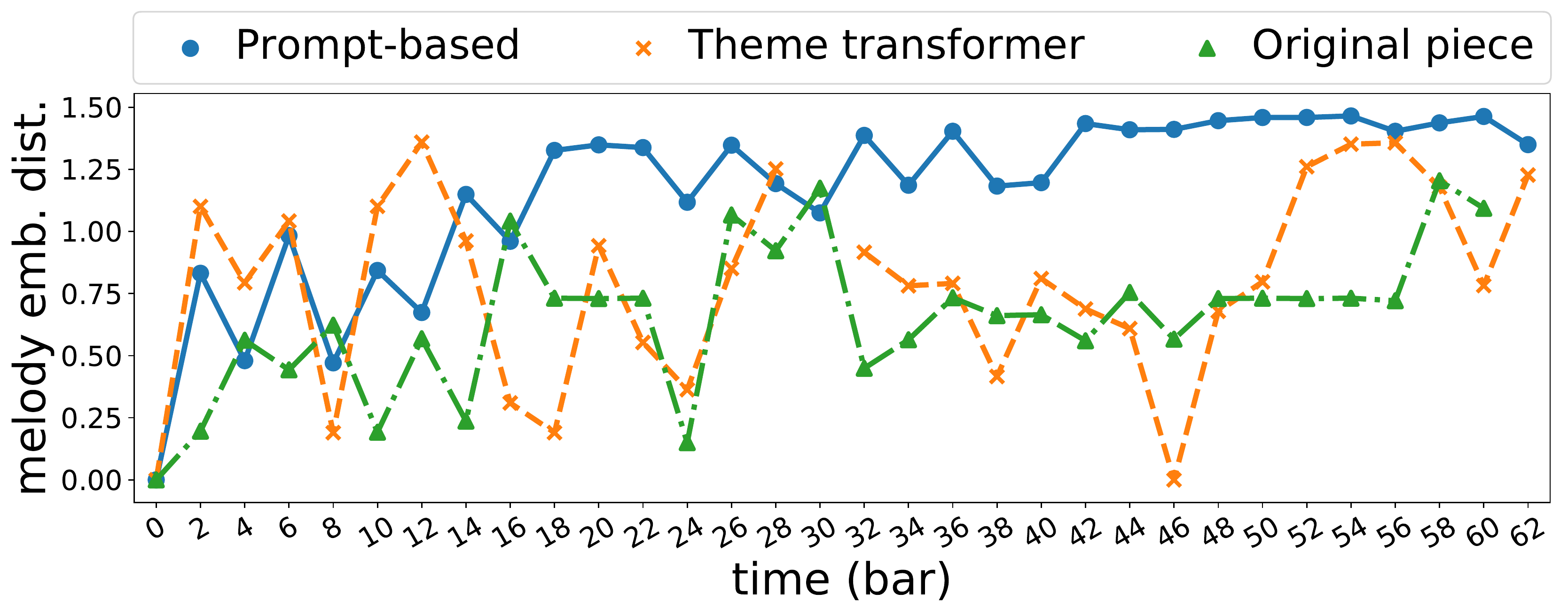}
\caption{Given either an original or generated piece, we split it into non-overlapping two-bar fragments, and evaluate the difference between each of these fragments with the 
\emph{beginning} (i.e., first) fragment of the same piece
using a distance measure computed in a ``melody embedding'' space (see Section \ref{sec:methodology:contrastive} for definition). The original piece here is an excerpt of the song `907.mid' from the test split of the POP909 dataset \cite{wang2020pop909}, starting from its thematic fragment. The other two are generated by a prompt-based Transformer \cite{huang2018music,huang20remitransformer} and the proposed theme-based Transformer, using that theme as the condition. We see that the melody drifts away from its beginning in the result of the prompt-based model (i.e., the blue line goes up). \iantext{Note that the embedding distance as the average of the whole POP909 dataset is 0.895.}}
\label{fig:the-influence-of-cond-over-time}
\end{figure}

\begin{figure*}
\centering
\includegraphics[width=0.94\textwidth,trim={0 0 0 3mm},clip]{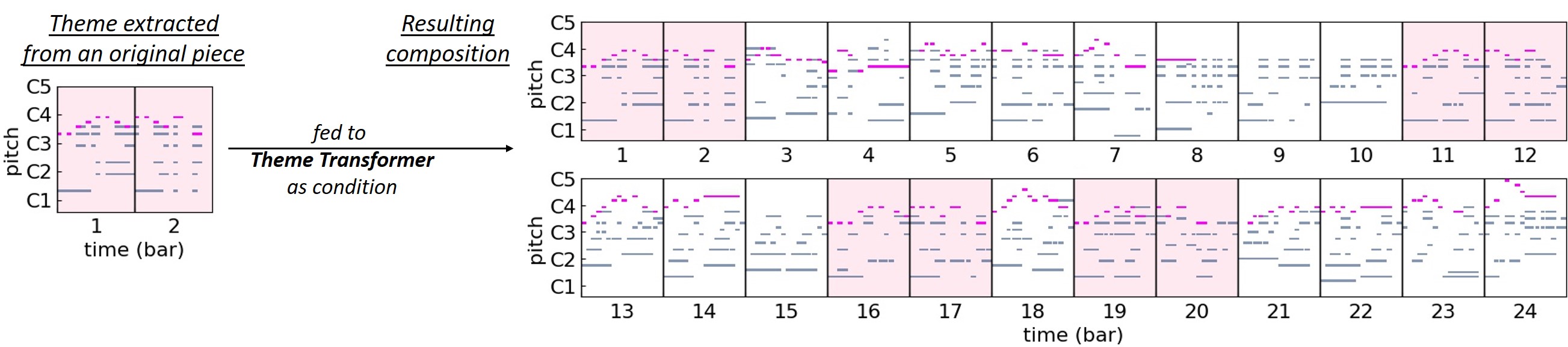}
\caption{The piano roll of the first 24 bars of a composition by Theme Transformer, conditioned on the theme of an unseen testing song `899.mid' from POP909. (Melody in magenta, accompaniment in grey, generated theme regions shaded in pink).}
\label{fig:composed-sample}
\end{figure*}

The self-attention mechanism allows us to learn a ``language model'' of music without hand-crafted rules and generate unique and diverse pieces that are internally consistent. 
However, by design, this strategy does not guarantee that the model would repeat or further develop the given condition (in the case of prompting), or the  previously generated material, in a meaningful way~\cite{wu20ismir}. 
Moreover, as the music piece is continuously being expanded, the influence of the beginning fragment of the piece might diminish over time, as demonstrated in Figure \ref{fig:the-influence-of-cond-over-time}.
Unlike human composers, even the state-of-the-art Transformer models for music generation do not learn to follow 
some central musical ideas
in their compositions, and they tend to lose the sense of a certain direction \cite{hernandezolivan21arxiv}.


To improve this shortcoming, we propose a novel conditional generation approach called ``theme-based'' generation in this paper. 
Similar to the popular ``prompt-based'' approach, the proposed approach also uses a music fragment (e.g., a musical theme) as input to condition the generative model.
However, there are two major conceptual differences.  First, we do not choose the conditioning fragment \emph{at random}; instead, we require that each of them to have \emph{multiple occurrences} in the corresponding music piece in our training set. 
Accordingly, we regard the conditioning fragment as a relevant thematic material for the piece.
Second, in addition to the self-attention mechanism, which governs memory to what has been previously generated, 
we establish a separate memory network that is dedicated to the provided condition. Namely, we demand the generative model to have \emph{two} sets of memory, rather than just one. The generation process now relies on the interplay of these two memory networks.

In this paper, we describe such a theme-conditioned Transformer 
that fulfills these two requirements with the following technical innovations. 
First, we present a clustering-based approach that groups together fragments of a music piece  that are close to one another in an embedding space learned by contrastive learning \cite{zhang21naacl} in an unsupervised way. 
Specifically, 
we employ a musically informed data augmentation strategy in contrastive learning
and use a density-based method for clustering  \cite{dbscan}. This strategy results in plausible variations among the fragments of the same cluster.
We then take the 
earliest-appearing fragment belonging to the largest cluster per training piece as the condition to train our generative model.

Second, instead of feeding the condition to a Transformer decoder as done in the prompt-based approach, we feed the condition to a separate Transformer \emph{encoder} that is trained jointly with the Transformer decoder. This way, we can use the cross-attention \cite{vaswani2017attention} between the encoder and decoder as the required secondary set of memory. The Transformer decoder, which serves as the generative model, not only self-attends to itself but also cross-attends to its encoder counterpart. 
Furthermore, we show that the vanilla sequence-to-sequence (seq2seq) encoder/decoder architecture \cite{huang2018music,bart,T-CVAE} is still not sufficient to enforce the influence of the condition when the generated music gets longer.
To address this issue, we propose a few modifications involving a novel gated parallel attention module, the use of theme-related tokens, and a theme-aligned positional encoding.



We present an empirical performance comparison between the proposed theme-conditioned Transformer and the conventional prompt-conditioned Transformer \cite{huang2018music,huang20remitransformer} for generating polyphonic piano music. Specifically, we train the models using the training split of the POP909 dataset \cite{wang2020pop909}, and conduct objective and subjective evaluations on its test split. 
We use the same clustering approach to find the most representative fragment from each testing song and use that as the condition at inference time.
In the objective evaluation, we use some general metrics from \cite{wu20ismir} and some novel theme-specific metrics proposed here.
The subjective evaluation entails an online listening test that involves listeners familiar and unfamiliar with the music pieces in the POP909 dataset. 
Our evaluation shows that the proposed Theme Transformer exhibits similar perceptual theme controllability and theme variation as original pieces in the test split of POP909.


For reproducibility, we provide all the original and generated music pieces used in our listening test on our demo website,\footnote{\url{https://atosystem.github.io/ThemeTransformer/}} which we encourage readers to listen to.
Figure \ref{fig:composed-sample} depicts an example of the generated music.
Furthermore, we provide all source code
at a public GitHub repository.\footnote{\url{https://github.com/atosystem/ThemeTransformer}}

\section{Related Work}

Unlike early rule- or knowledge-based systems \cite{jacob96,BozhanovMotif}, deep learning models do not use hand-crafted rules to generate music, 
thus allowing more flexibility in 
their output.
However, as music is an art of time, capturing the long-term dependency in music has been a major challenge for 
deep learning models. Roberts \emph{et al.} 
\cite{roberts2018hierarchical} presented a hierarchical latent vector model that uses a higher-level RNN to deal with bar-level
dependency of up to 16 bars long, and a lower-level RNN to generate notes in each bar independently. Huang \emph{et al.} 
\cite{huang2018music} greatly improved upon this by presenting the first Transformer decoder model for symbolic music generation, showing that a Transformer model can generate coherent minute-long polyphonic piano music with local repetitions and variations. 
Since then, many Transformer decoder-based models have been proposed. 

However, it remains difficult
for machine learning-based models, including Transformers, to generate pieces that exhibit clear musical phrases and sections with recurrence and development \cite{wu20ismir}. 
Efforts have been made to improve the overall repetitive structure of the music generated by either a Transformer model \cite{dai21ismir,zou2021melons}, 
an RNN model \cite{gabriele18ismir,jhamtani19ml4md},  
a Markov model \cite{velezdevilla21smc}, or an optimization-based approach \cite{8007229} in an \emph{unconditional} generation setting. Yet, none of them 
are designed for a \emph{conditional} (yet open-ended) generation setting where the generative model is asked to develop a given  
musical idea to generate a novel music piece.
The use of musical patterns to guide the generation process has been explored in early rule-based systems.
For example, Shan and Chiu \cite{ShanChiuAlgorithmicCompositionsMTA} used patterns automatically mined from an existing music corpus to generate a new composition from scratch; 
Zalkow \emph{et al.} \cite{ZalkowBrandGraf16StyleOptICMC} presented a rule-based system that modifies an existing composition 
to change its style.
Recent deep learning models (e.g., \cite{huang2018music, payne2019musenet,zou2021melons}) tend to use the first few notes of an existing piece as the priming sequence (i.e., the prompt) 
to trigger the generation process, but 
do not endow specific meanings upon them.
Deep generative models that explicitly consider musical patterns or ideas, to the best of our knowledge, have not been proposed before.

\section{Prompt-based vs. Theme-based Generation}

A Transformer decoder \cite{vaswani2017attention} 
learns the dependency among elements of a sequence by working on:
\begin{equation}
    p(x_t \vert x_{<t}) \,,
    \label{eq:uncond}
\end{equation}
where $x_t$ is the element of a sequence to be predicted at timestep $t$, and $x_{<t}$ 
is the subsequence consisting of all the preceding elements.
In each self-attention layer $l$, the model represents memory of the past by  two sets of hidden vectors, $[\mathbf{k}_1^{l},\dots,\mathbf{k}_{t-1}^{l}]$ and $[\mathbf{v}_1^{l},\dots,\mathbf{v}_{t-1}^{l}]$, a.k.a. the \emph{keys} and \emph{values}, respectively. 
In predicting $x_t$, the model (self-)\emph{attends} to each  preceding element by using the dot product between the \emph{query} hidden vector $\mathbf{q}_t^{l}$ at timestep $t$ and all the \emph{key} vectors $<t$ as the ``weights'' to linearly combine the \emph{value} vectors $<t$, yielding a hidden state $\mathbf{h}_t^{l}$ that is passed to the next layer as input.
Once a model is trained, the model can generate new sequences autoregressively, i.e., one element at a time based on the previously generated elements, in chronological order of the timesteps \cite{radford19language}. 

\textbf{Prompt-based conditioning} is arguably to date the most popular approach to condition the generation process of the Transformer decoder via a conditioning sequence \cite{fan18acl,keskar2019ctrl,brown2020language,fang2021transformer}.
The approach is straightforward: 
it 
regards the conditioning sequence $c_{1:\tau}=\{c_1,\dots,c_\tau\}$ as the
prefix sequence consisting of the first $\tau$ element
$\{x_1,\dots,x_\tau\}$ of the target sequence and lets the model generate a continuation starting from the timestep $\tau+1$. 
The conditioning sequence is thereby inherently part of the memory, affecting the subsequent elements by virtue of the generative modeling associated with Eq.~(\ref{eq:uncond}).

\textbf{Theme-based conditioning}, as proposed in this paper, employs a separate network (i.e., a Transformer encoder) to access the conditioning sequence in a different way. Therefore, the generation starts from the timestep $1$. 
The prediction of each $x_t$ is influenced by both the  decoder's \textit{self-attention} to its past (i.e., $x_{<t}$), and its \textit{cross-attention} to the keys and values in the encoder that represent $c_{1:\tau}$.
\begin{equation}
    p(x_t \vert x_{<t} ; c_{1:\tau}) \,.
    \label{eq:cond}
\end{equation}

In training a theme-based model, we need to have pairs that consist of a conditioning sequence $c_{1:\tau}$ and the target sequence $x_{1:T}$, where $T$ denotes the sequence length and naturally $\tau < T$. 
A na\"ive approach that uses a random subsequence from $x_{1:T}$ as $c_{1:\tau}$  does not work, because it is crucial to our application that the condition repeats itself a few times (with possible variations) in the target sequence. We introduce below how we find such pairs of $c_{1:\tau}$ and $x_{1:T}$.


\section{Finding Thematic Conditioning Materials}
\label{sec:theme-retrieval}

Creating theme-composition pairs that are \emph{musically} meaningful is challenging, because manual labeling the themes of music is a tedious task and requires significant musical expertise.
Automatic detection of motifs or themes also remains challenging, despite years of research~\cite{pinto10,VolkWK11ComputationalMusicologyICISO,lartillot14ismir,benamar17}.
The only public theme-related dataset,
the Musical Theme Dataset \cite{ZalkowBAM20MTDTISMIR}, is only for Western classical music and 
does not specify all positions where the themes occur in compositions.


Instead, we treat repeated music fragments \cite{janssen13CMMRa,LemstromSIA,ren16fma} 
as a plausible alternative thematic material.
Research has been done to automatically mine such patterns from a music corpus based on some mathematical assumptions  \cite{DavidMRECURSIA}. 
Most existing methods rely on hand-crafted rules and involve a large set of parameters that have to be manually tuned to achieve good performance.
After testing some of these methods in our pilot study, we found that they do not always lead to meaningful patterns.
Hence, we decided to develop a new approach that fulfils our own need.


\subsection{Sketch of our approach}
\label{sec:theme-retrieval:sketch}

Given a music piece $X$, our approach is to firstly partition the piece into several fragments, and then cluster the fragments in a 
learned embedding 
space. 
Each cluster contains similar, but not exactly the same, fragments of the same piece.
Then, we regard the \emph{largest} cluster as the ``theme cluster,'' and pick the fragment that appears the \emph{earliest} in the piece as the representative fragment of the whole piece, treating it as the thematic condition $c_{1:\tau}$. (The size of the theme cluster determines how many times the theme appears in the piece.) 
Subsequently, we can take any subsequence $x_{1:T}$ of $X$ 
and train a model that generates $x_{1:T}$ given the condition $c_{1:\tau}$.



We resort to \emph{contrastive learning} \cite{chen20icml,zhang21naacl} to build a \emph{retrieval} model of the thematic material. As detailed below, the contrastive learning approach allows us to specify which kinds of music fragments are supposed to be similar to the retrieval model, by using  synthetic pairs created with data augmentation. Furthermore, the retrieval model learns a representation $\mathbf{z} = \mathrm{Emb}(S)$ for each  fragment $S$, and accordingly a distance metric 
for evaluating the similarity of two fragments 
through a neural network, instead of using hand-crafted features. 


\subsection{Segmentation}
\label{sec:theme-retrieval:segmentation}

Partitioning a piece into fragments can be done by an automatic music segmentation algorithm \cite{miura09icmc,gttmimpl,GuanMelPhraseSeg}.
However, we found in our pilot study 
that the algorithms perform not well enough,
causing 
noises and over-segmentations.
We instead use a simple method to slice each piece into two-bar fragments without overlaps at the bar lines, while disregarding bars that do not contain melody.\footnote{Specifically, we first check the onset time of the first melody note in a piece. If it falls within the first-half of the corresponding bar, we start our two-bar slicing from that bar. Otherwise, we start the two-bar slicing from the next bar. Once the starting point is set, we continuously cut the melody into successive two-bar fragments until we reach a empty fragment. We then view the remaining sequence as a new piece and restart the procedure again till the end of the piece.} 
Empirically we found such a two-bar fragment carries sufficient information that allows the generation process of our Transformer to be influenced by the representative fragment.\footnote{According to our own listening of the original pieces in POP909, a music phrase in POP909 usually lasts for two bars, possibly due to the comfortable breath length for human singing. 
In addition, most of the phrases tend to start at the downbeat (i.e., the first beat) of a bar; none of them starts in the middle of a bar. 
Indeed, few of them start with an anacrusis (starting at the ``upbeat,'' the last beat in the previous bar). For those phrases, our segmentation method will not include the beginning notes.}

\subsection{Melody embedding learning by contrastive learning}
\label{sec:methodology:contrastive}

The next step is to learn a representation of each fragment. 
We propose to consider only the melody notes in a fragment for learning this representation, 
because themes and motifs are usually associated with the melody  \cite{OlivierMotivic}. 
In addition, focusing on the melody makes it easier to inform the retrieval model the variations of theme we expect to have by data augmentation, as explained below. 

We identify the melody notes for each  fragment using the
annotations of
POP909 dataset. (In cases where such annotations are not available, a melody extraction method such as the ``skyline'' algorithm \cite{uitdenbogerd99mm} may be used.) 
We then use an embedding network $\mathrm{Emb}(\cdot)$ to transform the melody of each fragment $S$ to a fixed-length embedding vector $\mathbf{z}$, called the \emph{melody embedding}.
The network uses a BERT-like architecture \cite{devlin18bert} comprising a number of bidirectional self-attention layers to compute the hidden state $\mathbf{h}$ of each note in the melody, and uses the mean of the hidden states from the last self-attention layer as the melody embedding $\mathbf{z}$.
To train the network, we use a \emph{contrastive head} 
that takes the melody embedding of two different fragments and computes their cosine similarity, $\mathrm{sim}(\cdot,\cdot)$. 
Given a batch of pre-defined \emph{positive} and \emph{negative} fragment pairs, the network learns by minimizing the following 
``contrastive loss'' \cite{chen20icml}:
\begin{equation}
    -\mathrm{log} \frac{\mathrm{exp}(\mathrm{sim}(\mathbf{z}_i,\mathbf{z}_j)/\alpha)}{\sum_{k} ~\boldsymbol{1}_{[k\neq i]}\mathrm{exp}(\mathrm{sim}(\mathbf{z}_i,\mathbf{z}_k)/\alpha)}\,,
    \label{eq:contrastive_loss}
\end{equation}
meaning that the model has to draw the embeddings of fragments in a positive pair (i.e., $(i,j)$) closer than those in a negative pair (i.e., $(i,k)$), where $\alpha$ denotes temperature.

A core idea of contrastive learning 
is to establish the positive and negative data pairs in an unsupervised way via ``data augmentation'' \cite{chen20icml}. In our case, this implies that we use rules to modify an input original fragment 
to create 
synthesized variations, and  regard the pair of the fragment and each of its variations as a positive data pair. Fragments from different pieces constitute the negative pairs. This way, the embedding model $\mathrm{Emb}(\cdot)$ learns to be \emph{invariant} to those variations employed in data augmentation. Specifically, we adopt the following three rules, motivated by common techniques in pop song writing:
\begin{itemize}
\item \emph{Pitch shift on scale}: Keep the same contour of a melody but shift them according to their positions in the musical scale. Note that it is different from uniformly shifting the pitches, because the distances between two adjacent notes are not equal in a Western heptatonic scale (i.e., there are whole steps and half steps).
\item \emph{Last note duration variation}: Randomly vary the duration of the last note in the melody.
\item \emph{Note splitting and combination}: Randomly pick a note in the melody and split it into two notes with the same pitch but half of the original duration each; or randomly combining two neighboring notes with the same pitch. 
\end{itemize}

\begin{figure}
\centering
\includegraphics[width=.22\textwidth]{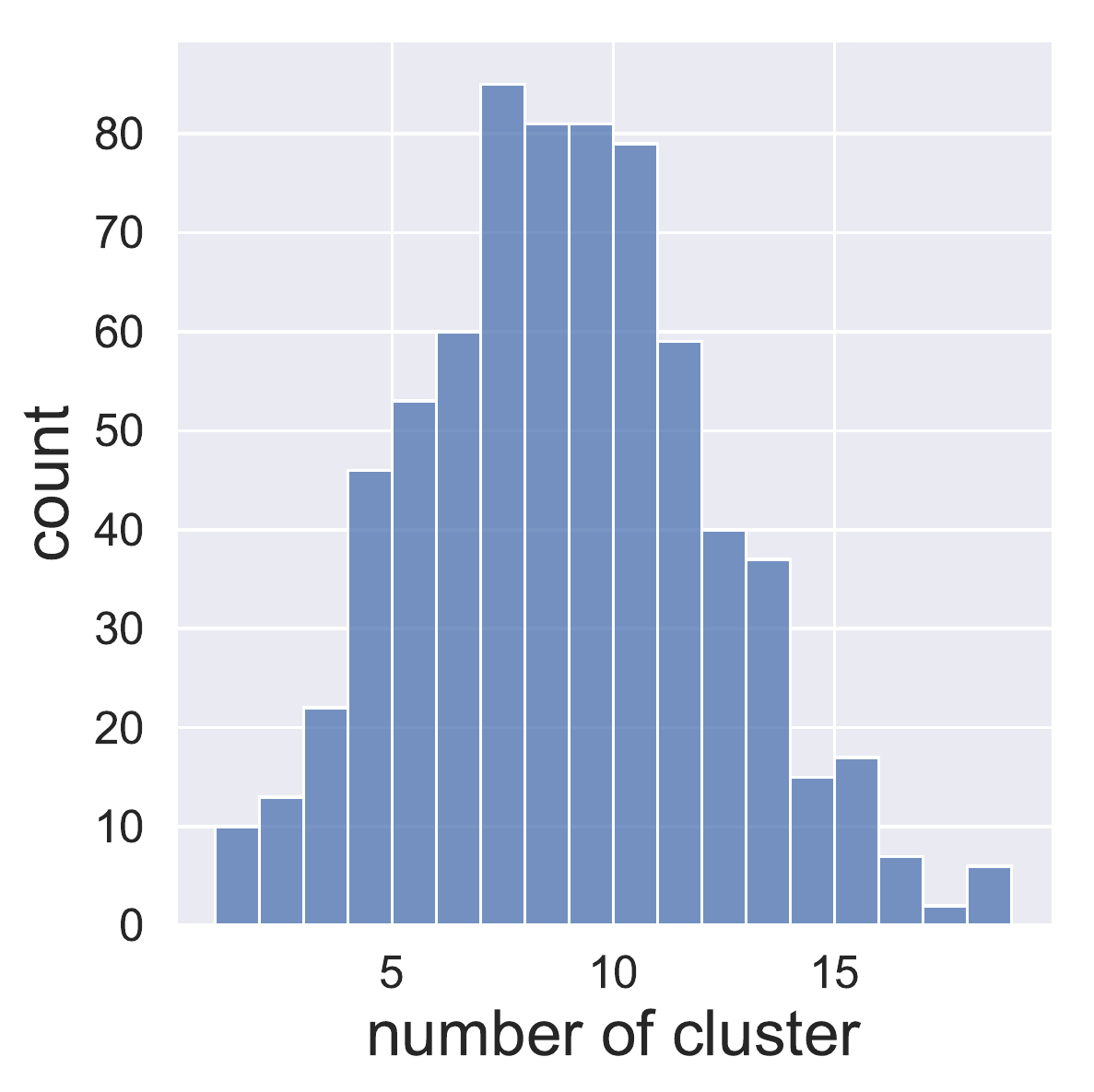}~~~~ 
\includegraphics[width=.22\textwidth]{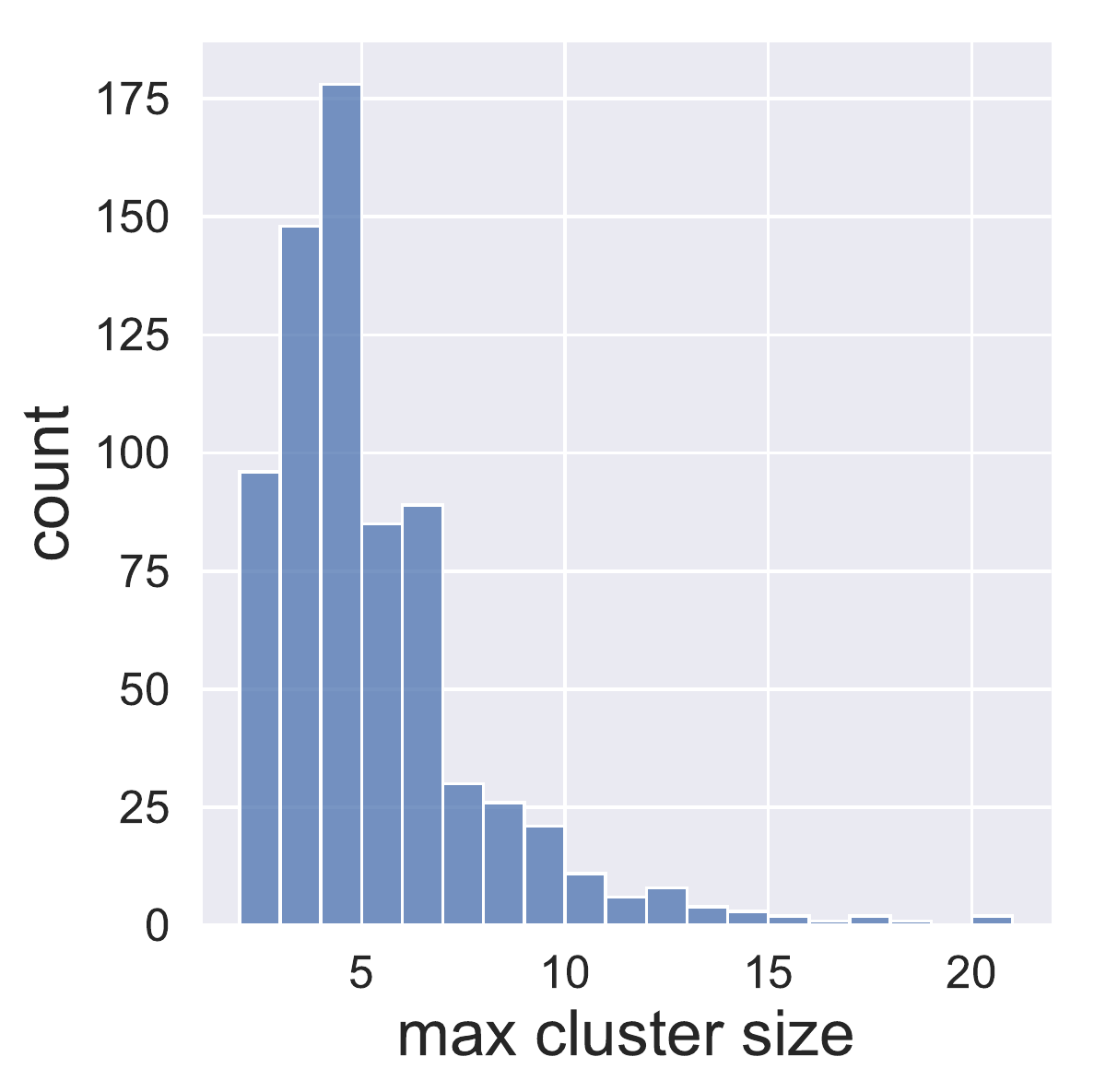} 
\caption{The histograms of (left) the number of clusters per song, and (right) the number of fragments in the largest cluster (i.e., the size of the theme cluster) per song.}
\label{fig:cluster_results}
\end{figure}

Because the embedding model is trained to discriminate similar and dissimilar fragments, we will also use it to set up objective metrics to evaluate the performance of our model:
\begin{equation}
    D(S_i,S_j) = \|\mathrm{Emb}(S_i) - \mathrm{Emb}(S_j)\|_2 \,.
    \label{eq:dist}
\end{equation}


In our implementation, the embedding model has a stack of 6 bidirectional self-attention layers,
followed by a contrastive head with an output dimension of 128.
The hidden dimension of the self-attention layers is set to 256. This model has in total 3M trainable parameters.
We train the embedding model using the contrastive loss described in Eq. (\ref{eq:contrastive_loss}) with the temperature hyperparameter $\alpha$ set to 0.5, using Adam as the optimizer and a batch size of 128 fragments, all of which are picked randomly from a different piece each time.
In each batch, we create two augmented versions for each fragment. 
The pair that consists of a fragment and one of its augmented version is treated as a positive pair. 
The other pairs are treated as negative pairs.
The training process converges after 2,760 training steps, landing on a final contrastive loss of 1.82.


\begin{figure*}[t]
\centering
\includegraphics[trim=0 18mm 0 0,clip,width=\textwidth]{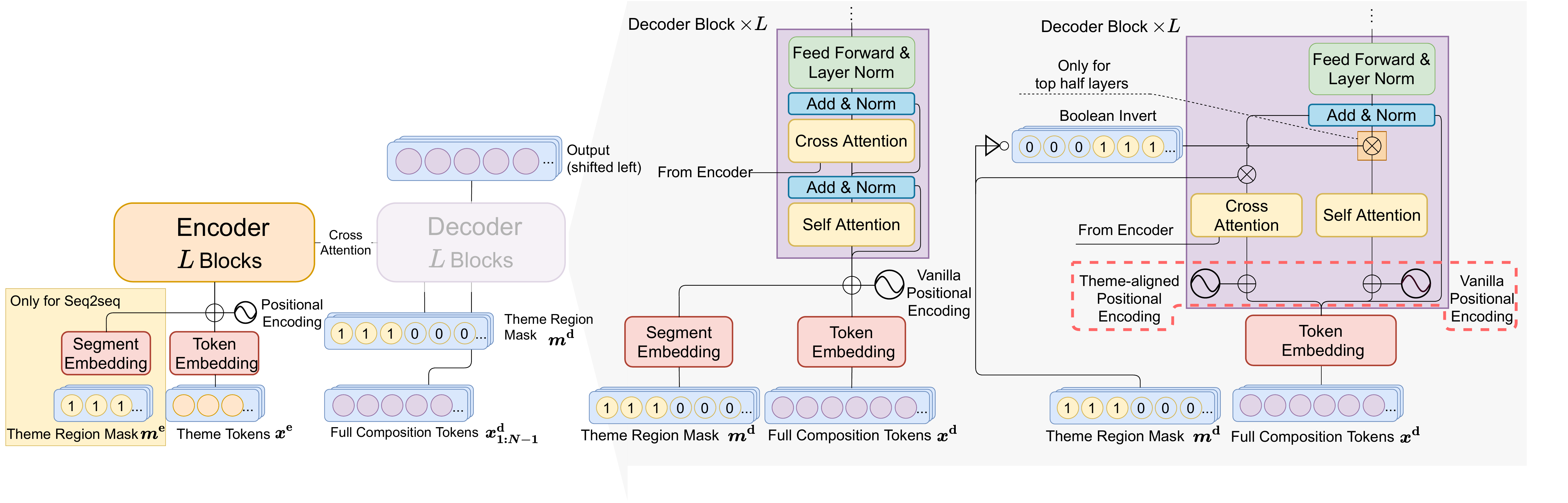} 
(a)~~~~~~~~~~~~~~~~~~~~~~~~~~~~~~~~~~~~~~~~~~~~~~~~~~~~~~~~~(b)~~~~~~~~~~~~~~~~~~~~~~~~~~~~~~~~~~~~~~~~~~~~~~~~~~~~~~~~~(c)
\caption{(a) Schematic overview of the proposed Transformer architecture for theme-conditioned music generation. Diagrams of the decoder architecture for (b) the basic seq2seq model that utilizes segment embedding ($\mathrm{SE}$), and (c) the proposed Theme Transformer that utilizes parallel attention modules with XOR gating (the two ``$\otimes$''s) and separate positional encodings ($\mathrm{PE}$s).}
\label{fig:models}
\end{figure*}

\subsection{Clustering} 
\label{sec:theme-retrieval:clustering}

We use a density-based  algorithm called DBSCAN \cite{dbscan} to cluster the fragments of a piece. 
For each piece, DBSCAN forms a variable number of clusters by requiring the distance between the cluster centroid and every fragment included by the cluster to be smaller than a pre-defined threshold $\epsilon$, in the learned melody embedding space.
In this way, we can control the closeness of fragments in a cluster, and at the same time adaptively decide the number of clusters of a piece according to the repetitive nature of the piece.
In practice, 
the DBSCAN algorithm has two user-defined hyperparameters: the minimal number of components (i.e., fragments in our case) that constitute a cluster, and the maximal distance $\epsilon$  between components in a cluster to the cluster centroid \cite{dbscan}.
We set the former to 2.
We set the latter according to our empirical observation of the distance between the embeddings of some toy sequences with perceptually tolerable alternations we fed to the embedding model (when the model converges). 
We finally set $\epsilon$ to 0.13, for it allows DBSCAN to include those similar sequences to the same cluster.\footnote{Accordingly, it seems reasonable, as shown later in Table \ref{tab:objective_result}, that the average \emph{theme inconsistency} and \emph{theme uncontrollability} values (which are also calculated by $D(\cdot,\cdot)$) of the original pieces in the testing data turn out to be below the value of $\epsilon$.}

Figure~\ref{fig:cluster_results} shows the histograms of the number of resulting clusters per song, and the  the number of fragments in the theme cluster.
Moreover, we provide examples of fragments belonging to the same cluster on our demo website\footnote{\url{https://atosystem.github.io/suppl_clusterMID.zip}} in MIDI format. We note that DBSCAN would discard fragments of a piece that do not not belong to any clusters.

Note that, the melody of a fragment is only used for clustering. As a conditioning fragment is an excerpt of our music pieces, it contains both the melody and its accompaniment.

\section{Theme-Conditioned Transformer}

\subsection{\iantext{Seq2seq Model}}

A theme-conditioned Transformer can be implemented by a seq2seq Transformer \cite{huang2018music,T-CVAE}, with the \emph{source sequence} (i.e., the encoder input) being a condition $c_{1:\tau}$
and the \emph{target sequence} (i.e., the output of the decoder) being a segment $x_{1:T}$ of the corresponding music piece.
The encoder comprises several bidirectional self-attention layers, and the decoder uses a
\emph{cascade} of uni-directional self-attention and cross-attention modules in each layer (see Figure~\ref{fig:models}(b)) to integrate information from both the decoder itself and encoder.

In Transformer-based music generation, a music piece is usually represented as a sequence of ``event tokens,'' such as \textsc{Note-On} and \textsc{Note-Duration} \cite{huang20remitransformer}.
Using the theme occurrences obtained with our contrastive clustering method, 
we propose to add \textsc{Theme-Start} and \textsc{Theme-End} tokens before and after each two-bar fragment belonging to the theme cluster.
These tokens divide a token sequence into \emph{theme regions} and \emph{non-theme regions}.
Tokens within a theme region are largely affected by the thematic conditioning material.
The condition can also influence tokens in a non-theme region without explicitly manifesting itself.%

Segment embeddings are commonly used to help a Transformer distinguish different parts of a sequence, e.g., \textit{question} and \textit{answer}  \cite{devlin18bert}.
We can use them to highlight the difference between the theme and non-theme regions in self-attention. 
As Figure \ref{fig:models}(a) shows, from our  clustering result, we  feed to the decoder a binary \textit{theme region mask} $\mathbf{m}^\mathrm{d}=\{m^\mathrm{d}_1,\dots,m^\mathrm{d}_T\}$,
where $m^\mathrm{d}_t = 1$ indicates that the timestep $t$ is within a theme region, and $m^\mathrm{d}_t = 0$ otherwise.
The encoder 
is given an all-$1$'s theme region mask $\mathbf{m}^\mathrm{e}=\mathrm{1}_\tau$.
Then, $\mathbf{m}^\mathrm{e}$ and $\mathbf{m}^\mathrm{d}$ pass a shared segment embedding ($\mathrm{SE}$) to become part of the input to the first self-attention layer of the encoder and decoder.
Specifically, the input vector  $\mathbf{\tilde{x}}_t^\mathrm{d}$  of the decoder at timestep $t$ is
$\mathbf{\tilde{x}}_t^\mathrm{d}= \mathrm{TE}(x^\mathrm{d}_t) + \mathrm{PE}(t) + \mathrm{SE}(m^\mathrm{d}_t)$,
where
$\mathrm{TE}$ is the token embedding also shared by the encoder and decoder, and $\mathrm{PE}$ the vanilla \emph{sinusoidal} positional encoding  
\cite{vaswani2017attention,ke2021rethinking,wang2021position}.
The input to the encoder
(i.e., $\mathbf{\tilde{x}}_t^\mathrm{e}$) is obtained similarly.

\subsection{Proposed Theme Transformer}
However, unlike many seq2seq tasks in natural language processing (NLP), where the target sequences tend to be short (e.g., in question answering \cite{devlin18bert}, machine translation \cite{cho-etal-2014-properties}, or summarization \cite{yangBertSum}), in our music generation task the length of the target sequences may well be longer than 64 bars, which translate to roughly 4,000 tokens (elements) 
for our dataset.
Using the vanilla positional encoding ($\mathrm{PE}$)  
to keep track of the distance (in terms of timesteps) 
between tokens
may greatly attenuate the influence of the condition at later timesteps of the decoder, i.e., when $\tau \ll t$.
Moreover, the relationship between the source and target sequences may be subtler in our task compared to the NLP tasks.
Due to these issues, the decoder may ignore the encoder entirely. We propose the following modifications to avoid this problem. 






\subsubsection{Gated parallel attention}
To equalize the importance of  self- and cross-attention,
we adopt the idea of Rashkin \emph{et al.} \cite{Hannah2020parallel} and place them \emph{in parallel} (instead of cascaded), as depicted in Figure \ref{fig:models}(c). 
Moreover, we propose a novel \emph{XOR gating} mechanism, such that the decoder \emph{only} uses self-attention when the current timestep $t$ falls within a non-theme region (i.e., $m_t^\mathrm{d}=0$) in all its $L$ layers, and \emph{only} uses cross-attention when $t$ falls within a theme region\footnote{At inference time, this implies that there is already a \textsc{Theme-Start} token but not yet a \textsc{Theme-End} token in the recent past.}  (i.e., $m_t^\mathrm{d}=1$) in \emph{some} of its $L$ layers. 
Specifically, within a theme region, we only turn off self-attention 
in the \textit{top half} layers, so that self-attention still takes effect and contributes to the local continuity of music. Mathematically, the output of the $l$-th decoder layer at $t$, denoted as $\mathbf{h}_t^l$, can be written as:
\begin{equation}
    \mathbf{h}_t^l =\begin{cases}
    \mathit{m_t}\,\mathbf{h}^{l,(\mathrm{cross})}_t + (1-\mathit{m_t})\,\mathbf{h}^{l,(\mathrm{self})}_t\,, & l > L/2 \\
    \mathit{m_t}\,\mathbf{h}^{l,(\mathrm{cross})}_t +\mathbf{h}^{l,(\mathrm{self})}_t\,, & l\leq L/2 
    \end{cases} 
\end{equation}
where 
$\mathbf{h}^{l,(\mathrm{cross})}_t$ and $\mathbf{h}^{l,(\mathrm{self})}_t$ are the outputs of cross- and self-attention, respectively.
For the self-attention route, we have tried other arrangements of the XOR gates (i.e., not placed in the top half), finding no notable differences.
\cite{vaswani2017attention}

\subsubsection{Theme-aligned positional encoding}
Finally, instead of using only one set of $\mathrm{PE}$ as shown in Figure \ref{fig:models}(b), we propose a novel idea that uses \emph{separate} $\mathrm{PE}$s for the self- and cross-attention routes in parallel attention, as depicted in Figure \ref{fig:models}(c).
Our goal is to \emph{align} the $\mathrm{PE}$ 
of the theme regions on the decoder side with the $\mathrm{PE}$ of the condition for the cross-attention route.
For example, let's assume that the decoder generates a \textsc{Theme-Start} token at timestep $\kappa$, where $\tau \ll \kappa$. The decoder may not easily cross-attend to any tokens within $c_{1:\tau}$ when predicting $x_{\kappa+1}$, because $\mathrm{PE}(\kappa+1)$ would be greatly different from any $\mathrm{PE}(\rho)$, for $\rho\in[1,\tau]$, under the vanilla sinusoidal $\mathrm{PE}$ \cite{vaswani2017attention} due to the big gap between $\tau$ and $\kappa+1$. 
To remedy this, we ``reset the clock'' and use $\mathrm{PE}(1)$ as the positional encoding for timestep $\kappa+1$ when calculating the cross-attention. Similarly, we use $\mathrm{PE}(t)$ as the positional encoding for timestep $\kappa+t$ when calculating the cross-attention, until the decoder generates a \textsc{Theme-End} token.
In our implementation, we inject $\mathrm{PE}$s 
in \textit{every} layer, instead of just once at the input layer as conventionally done.


It can be seen that the use of separate $\mathrm{PE}$s goes naturally in tandem with parallel attention, rather than cascaded attention.
In the proposed Theme Transformer, we drop the $\mathrm{SE}$ entirely and uses the parallel attention with XOR gating and separate $\mathrm{PE}$s in the decoder.

\section{Implementation Details}

\subsection{Dataset}
We train our models using the piano covers of Mandarin pop music from the POP909 dataset \cite{wang2020pop909}, 
which is composed of the piano covers of 909 Mandarin pop songs originally composed by 462 artists, released from the earliest in 1950s to the latest around 2010 \cite{wang2020pop909} and the piano arrangements in these covers were created by professional musicians.
Each arrangement is stored in three separate MIDI tracks: MELODY (transcription of the lead melody, which corresponds to the vocal), BRIDGE (the secondary melodies or
lead instruments), and PIANO (the main body of the accompaniment), making it trivial to identify the melody.
In our work, we omit the BRIDGE track, since its functionality is in between melody and accompaniment.
Besides, perceptually the quality of the original tracks does not degrade much without the BRIDGE track.
Following \cite{hsiao21aaai}, we pick only the songs with a 4/4 time signature and quantize the note onset times and duration to be multiples of 1/4 beat (ignoring triplets) for simplicity, using the beat annotations provided by POP909. 
In addition, we filter out songs that change their key within the songs to avoid dealing with such cases when applying data augmentation in melody contrastive learning.
This leads to a final collection of 713 songs.
We reserve 4\% of them (i.e., 29 songs) for testing and the others for model training.



\begin{table}[t]
\centering
\begin{tabular}{|l|r|r|} 
\hline
Token type & Piano Representation & Melody Representation \\ 
\hline
\textsc{Note-Pitch} & 127$\times$2 & 127\\ 
\textsc{Note-Duration} & 64$\times$2 & 64\\ 
\textsc{Note-Velocity} & 126$\times$2 &---\\ 
\textsc{Tempo} & 76 &---\\ 
\textsc{Bar} & 1 &---\\ 
\textsc{Subbeat} & 16 &---\\ 
\textsc{Theme} & 2 &---\\ 
\textsc{Padding} & 1 &1\\ 
\hline
Total & 730 & 192\\ 
\hline
\end{tabular}
\caption{Vocabulary of our token representation, which is used for music generation and contrastive clustering. Note that, for the prompt-based baseline, we do not use the theme tokens.}
\label{tab:vocab}
\end{table}

\subsection{Token representation}

We consider three types of tokens to represent each song from POP909 by a token sequence.
First, each note in a song is described by three \emph{note-related} tokens: 
\textsc{Note-Pitch} (from \textsc{D0} to \textsc{G9}), 
\textsc{Note-Duration} (ranges from 1/4 beat to 16 beats, in multiples of 1/4 beat), 
and \textsc{Note-Velocity} (how hard the note is played; from 1 to 126). 
Moreover, we adopt a multi-track representation \cite{payne2019musenet,donahue2019lakhnes} and use separate note-related tokens for melody and accompaniment. 
For example, to describe the pitch of a melody note in C5, we use the token \textsc{Note-Pitch:Melody:C5}. 
Second, following \cite{huang20remitransformer},
we use the following \emph{metric-related} tokens to represent the advance in time. 
\textsc{Bar} marks the beginning of a new bar, which is uniformly divided into 16 subbeats. 
\textsc{Subbeat} indicates the onset time of a note among these 16 possible positions within a bar.
\textsc{Tempo} describes the speed of a song at the beat-level, and takes values from 17 to 194 BPM, with an interval of 3. 
Lastly, we have \emph{theme-related} tokens: \textsc{Theme-Start} and \textsc{Theme-End}.
(Following the convention in prompt-based models, we do not use the two theme tokens 
in the prompt-based baseline.)
In total, our vocabulary for the piano contains 730 unique tokens.\footnote{Because we use 512-token sequences as input, rather than the whole pieces, We do not have \textsc{EOS} (i.e., end-of-song) tokens \cite{hsiao21aaai}. Hence, our models generate endless pieces following a given condition at inference time. In our evaluation, we truncate them at the 64-th bar.}
The songs in our dataset have $\sim$95 
bars on average, which translate to 5,249 tokens per song on average using the piano representation.
Accordingly, a 512-token sequence employed in model training (i.e., $x_{1:T}$) contains 
about nine 
bars on average. A  two-bar conditioning fragments (i.e., $c_{1:\tau}$) contains 121.9 tokens on average.

While the piano representation is used in the Transformer models for generation, we need a separate melody representation to be used in the contrastive-learning based retrieval model for finding thematic conditioning material.
For the melody representation, we add a REST token as another instance of \textsc{Note-Pitch}. In this way, the ending a note would be exactly the beginning of a new note (including REST), as the melody is monophonic. Accordingly, we can faithfully represent a melody without metric-related tokens. In our dataset, the average length of such a melody sequence, which is fed to the embedding model, is 24.3 tokens.\footnote{We note that, the pieces generated by our Transformer models are sequences of the piano tokens using the vocabulary on the left hand side of Table \ref{tab:vocab}. To get the melody embedding of two-bar fragments of the generated pieces for computing the objective metrics \emph{theme inconsistency} and \emph{theme uncontrollability} (cf. Section \ref{sec:obj:metrics}), we need to ``translate'' the tokens to what described on the right hand side of Table \ref{tab:vocab}.
In doing so, we  pick the melody notes in the generated pieces, add REST tokens when appropriate, and then discard tokens other than pitch- and duration-related ones.}

Table \ref{tab:vocab} lists the token types and the number of unique tokens for each type in our token representation of the piano and the melody, respectively.

\subsection{Model settings}

We use a 6-layer encoder and a 6-layer decoder (i.e., $L=6$) for both the basic \iantext{\textbf{seq2seq Transformer}} and the \iantext{\textbf{proposed Theme Transformer}}. Both are theme-conditioned. 
In addition, we implement a \iantext{\textbf{baseline}} prompt-conditioned Transformer decoder \cite{huang2018music,huang20remitransformer} that also has 6 layers. 
All the models are trained with the Adam optimizer ($\beta_1=.9, \beta_2=.99$) \cite{kingma2014adam} and teacher forcing
to minimize the training negative log-likelihood, $-\sum_{t=1}^T \log p(x_t \, | \, x_{<t})$, where $T=512$,
using $2\times10^{-4}$ learning rate
and a batch size of 8. 




All our models have 8 heads for multi-head attention, 256 hidden dimensions, 1,024-dim feed-forward layers, and GeLU as the activation function. The total number of trainable parameters is 11M for the basic seq2seq model and the proposed Theme Transformer, and is 5M for the prompt-based decoder-only baseline.
The training NLL of all these three models drops below 0.55 in roughly two days when being trained separately on an NVIDIA V100 GPU.
While the theme regions in the training data are all two-bar long, it turns out that the theme-based models also learn to have each generated theme region to be in two bars.

For the training data, to ensure that the parameters associated with cross-attention can be updated every training step, we select the 512-token sequences including either a \textsc{Theme-Start} or \textsc{Theme-End} for the two theme-based models, ending up having 2,529 sequences for training.
However, for the prompt-based model, theme-related tokens are not used at all, 
so we follow the convention in the literature and pick 2,529 random subsequences of the training pieces for training  the prompt-based model.



In the inference phase, the prompt-based model uses the theme condition as the prefix of the sequence to be generated. Therefore, the generated piece of the prompt-based model  always start from a theme region. 
However, this is not the case for the theme-based models, because the models may not choose to exactly copy the theme condition in the beginning of its generation. Such a difference is fine in practical applications, but it may render the comparison between the generated pieces in the listening test inconvenient or difficult. Therefore, in our implementation, we use a \textsc{Theme-Start} token as the prefix of the sequence to be generated for the theme-based models at inference time.


\begin{table*}[t]
\centering
\small

\begin{tabular}{|l|ccc|ccc|} 
\hline\
& Pitch class & Melody & Grooving  & Theme incon- & Theme uncon- & Theme gap \\ 
&consistency$\uparrow$&inconsistency$\downarrow$&consistency$\uparrow$&sistency$\downarrow$&trollability$\downarrow$& (in \# bars)\\
\hline \hline

\iantext{Baseline (prompt-based)} \cite{huang2018music,huang20remitransformer}  
&
.59$\pm$.07&
.33$\pm$.38&
.84$\pm$.09& 
---&---&---\\ 
\iantext{Seq2seq Transformer} \cite{huang2018music}&
{.61$\pm$.04}& 
.46$\pm$.28&
.90$\pm$.06& 
1.01$\pm$0.05&
1.10$\pm$0.14&
6.02$\pm$1.91\\
Theme Transformer (proposed)&
{.61$\pm$.06}&
{.13$\pm$.24}& 
{.92$\pm$.07}&
{0.27$\pm$0.26}& 
{0.24$\pm$0.20}& 
{9.48$\pm$3.59}\\ 
\hline
Original pieces 
& 
.65$\pm$.05& 
.09$\pm$.18&
.74$\pm$.10& 
0.05$\pm$0.05& 
0.04$\pm$0.04& 
12.24$\pm$11.32\\
\hline
\end{tabular}
\caption{Result of the objective evaluation. The scores are in general closer to those of the original pieces (i.e., test split of POP909) the better. We do not use  theme-related metrics for the prompt-based baseline, for it does not generate theme tokens.}
\label{tab:objective_result}
\end{table*}

\begin{table*}
\centering
\small
\begin{tabular}{|lrrcc|cccc|} 
\hline
 &sequence & \#self-attn&\multirow{2}{*}{$\mathrm{SE}$}&separate& Melody & Theme incon- & Theme uncon- & Theme gap \\ 
&length $N$&layers $L$&&$\mathrm{PE}$s& inconsistency$\downarrow$&sistency$\downarrow$&trollability$\downarrow$& (in \# bars)\\
\hline \hline
\multirow{3}{*}{Theme Transformer} & 512 & 6 & & $\surd$ & 
.13$\pm$.24 &
.27$\pm$.26 &
0.24$\pm$0.20 &
~9.48$\pm$3.59\\
&1,024& 6 & & $\surd$ & 
.07$\pm$.15  &
.27$\pm$.21 &
0.26$\pm$0.19 &
13.70$\pm$8.34\\
& 512 & 6 & $\surd$ &  & 
.19$\pm$.23  &
.55$\pm$.27 &
1.07$\pm$0.26 &
~7.70$\pm$3.68 \\
\hline
\multirow{2}{*}{\iantext{Baseline} \cite{huang2018music,huang20remitransformer}}
&512&6&&& .33$\pm$.38&---&---&---\\
&512&12&&&.33$\pm$.38  &---&---&---\\
\hline\hline
Original pieces 
&&&&& 
.09$\pm$.18&
.05$\pm$.05& 
0.04$\pm$0.04& 
12.24$\pm$11.32 \\ 
\hline
\end{tabular}
\caption{Result of an ablation study that evaluates variants of the implemented models }
\label{tab:ablation}
\end{table*}

\section{Objective Evaluation}

\subsection{Evaluation Setup and Metrics}
\label{sec:obj:metrics}
In the objective evaluation, we let each model generate 64-bar polyphonic piano music using the thematic conditions retrieved from each of the 29 testing songs of POP909.
We evaluate the result using two sets of metrics. The first set concerns with the generated music itself, probing the general coherence within a piece (which conventional prompt-based Transformer is known to be good at \cite{huang2018music}). The second set examines the relationship between each generated piece with the corresponding condition, studying how the condition manifests itself in the generated piece.

The first set has three metrics, each concerning a different aspect.
The first metric uses the overlapping area \cite{yang18evaluation} of the pitch class histogram of pairs of bars in a generated music to assess \emph{pitch class consistency}.  
The second one, which is novel, uses Eq.~(\ref{eq:dist}) to measure \emph{melody inconsistency} by comparing the melody embeddings $\mathbf{z}$ of each two-bar fragment in the latter 32 bars of a piece to the melody embedding of the beginning (i.e., first) two bars $S_1$ of the same piece, considering only the melody notes.
We pick the fragment $S_*$ with the closest distance to $S_1$ and report $D(S_1,S_*)$.
The last metric, \emph{grooving consistency}, as proposed by  \cite{wu20ismir}, studies the coherence in rhythm. 

The second set of metrics are all theme-related and are newly proposed here. 
We develop \emph{theme inconsistency} to evaluate the average difference in melody embedding among all pairs of different theme regions (delimited by \textsc{Theme-Start} and \textsc{Theme-End} tokens) in a generated piece, i.e.,
$\frac{2}{N(N-1)}\sum_{i,j} D(\Gamma_i,\Gamma_j)$,
where $\Gamma_i$ denotes a theme region and $N$ the number of theme regions in a  piece. 
By \emph{theme uncontrollability}, we evaluate the closeness between the theme regions and the condition,
$\frac{1}{N}\sum_{i=1}^N D(c_{1:\tau},\Gamma_i)$.
Lastly, \emph{theme gap} calculates the length (in \# bars) between two successive \textsc{Theme-Start} tokens.

\subsection{Main Result}
\label{sec:obj:main}
Table~\ref{tab:objective_result} displays the average result across the testing songs. We make the following observations.
First, all the models perform similarly on \emph{pitch class consistency}, suggesting that they all learn to adhere to a certain key in their generations. 
Second, the Theme Transformer performs much better than the other two in \emph{melody consistency}, implying that its generation follows a certain theme better.
Third, compared to the original pieces, all the models have overly high \emph{grooving consistency}, suggesting a lack in diversity in the generated rhythm.

The result on the theme-related metrics shows that the Theme Transformer performs much better than the basic seq2seq Transformer model in harnessing the condition in the generation, suggesting the effectiveness of the proposed theme-aligned positional encoding and gated parallel attention.
Yet, there is still a large gap between the Theme Transformer and original piece in  theme uncontrollability and theme gap, providing room for future work. 


The result in theme gap suggests that our model has difficulty learning the timing to reuse a theme. This might be due to our short training sequence length (i.e., 512 tokens). 
The model cannot learn long-term relation of theme occurrences 
at different song sections.

%
 
\begin{figure*}
\begin{tabular}{cc} 
\includegraphics[width=.45\textwidth]{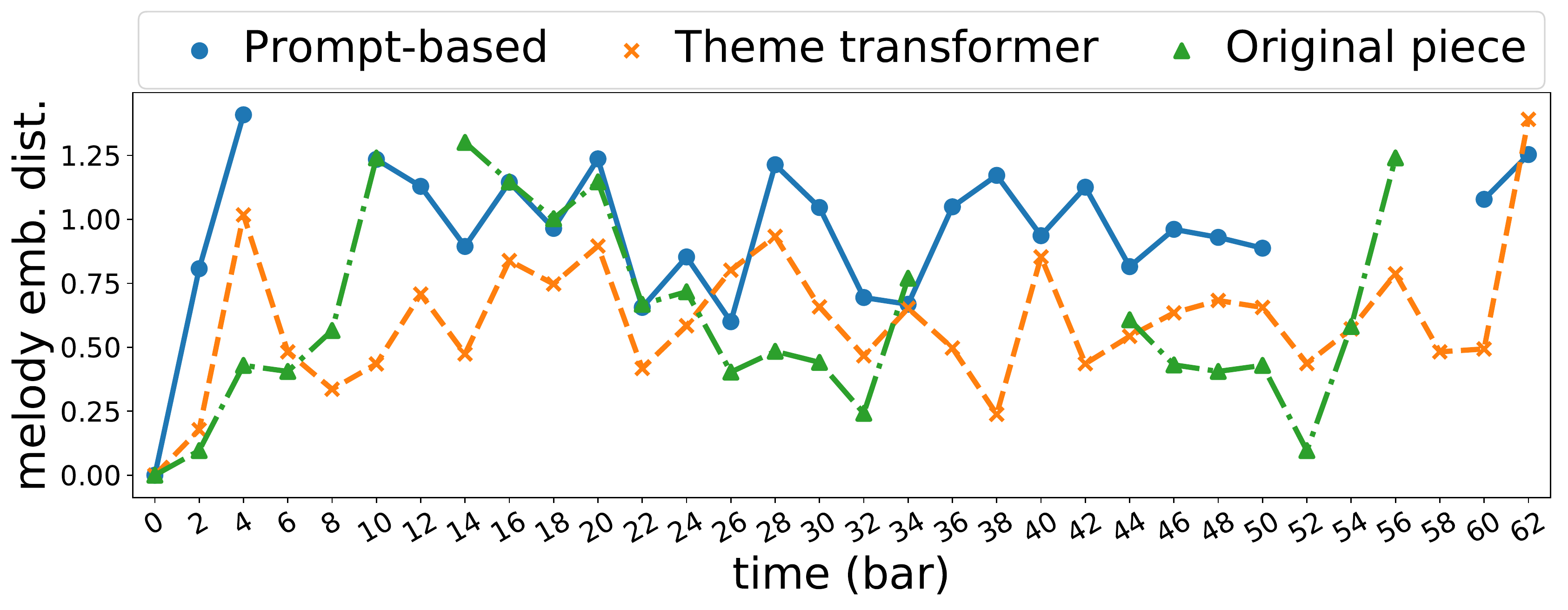}
&
\includegraphics[width=.45\textwidth]{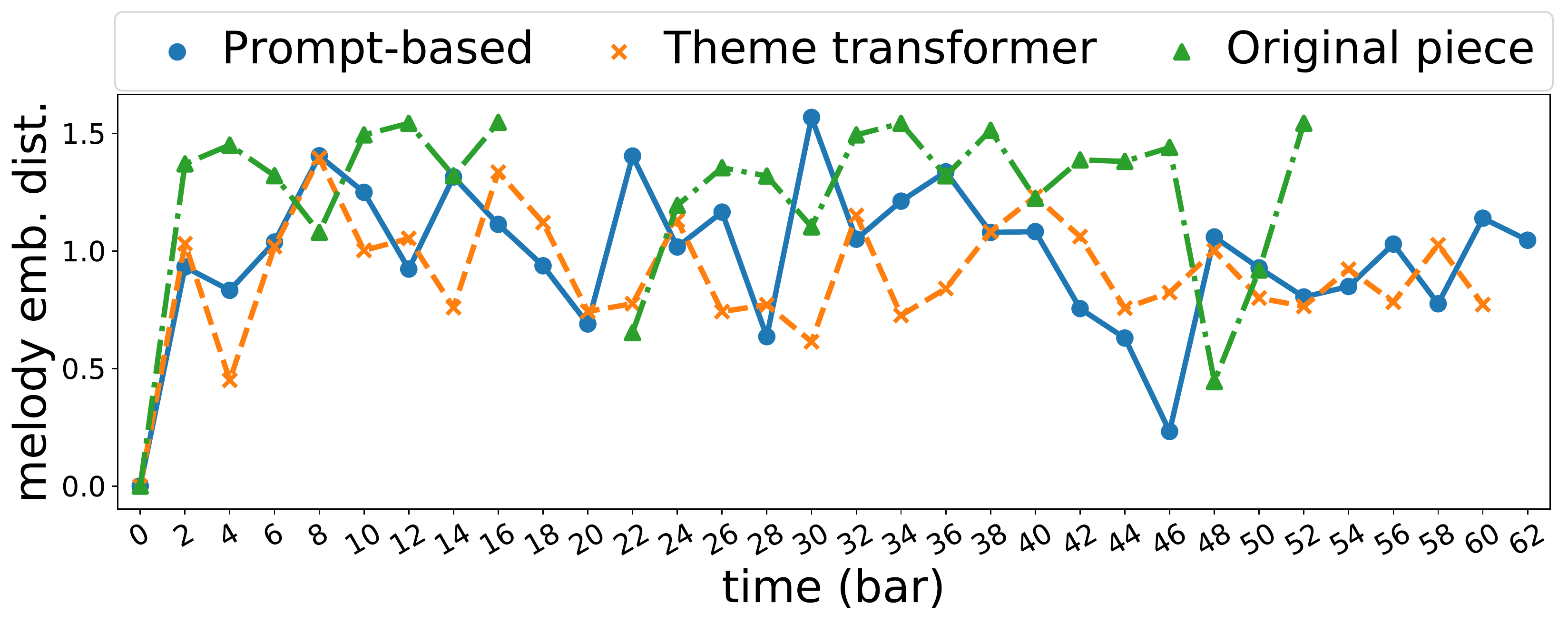} 
\\
\includegraphics[width=.45\textwidth]{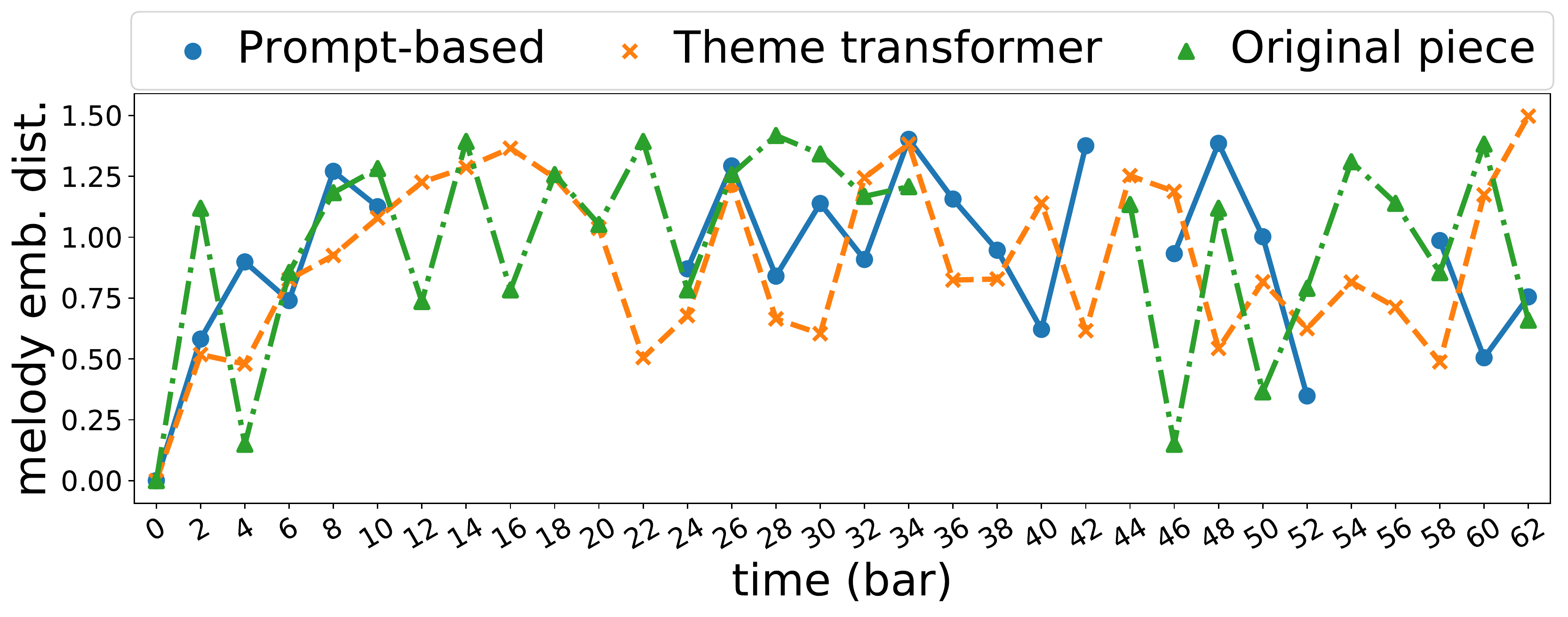}
&
\includegraphics[width=.45\textwidth]{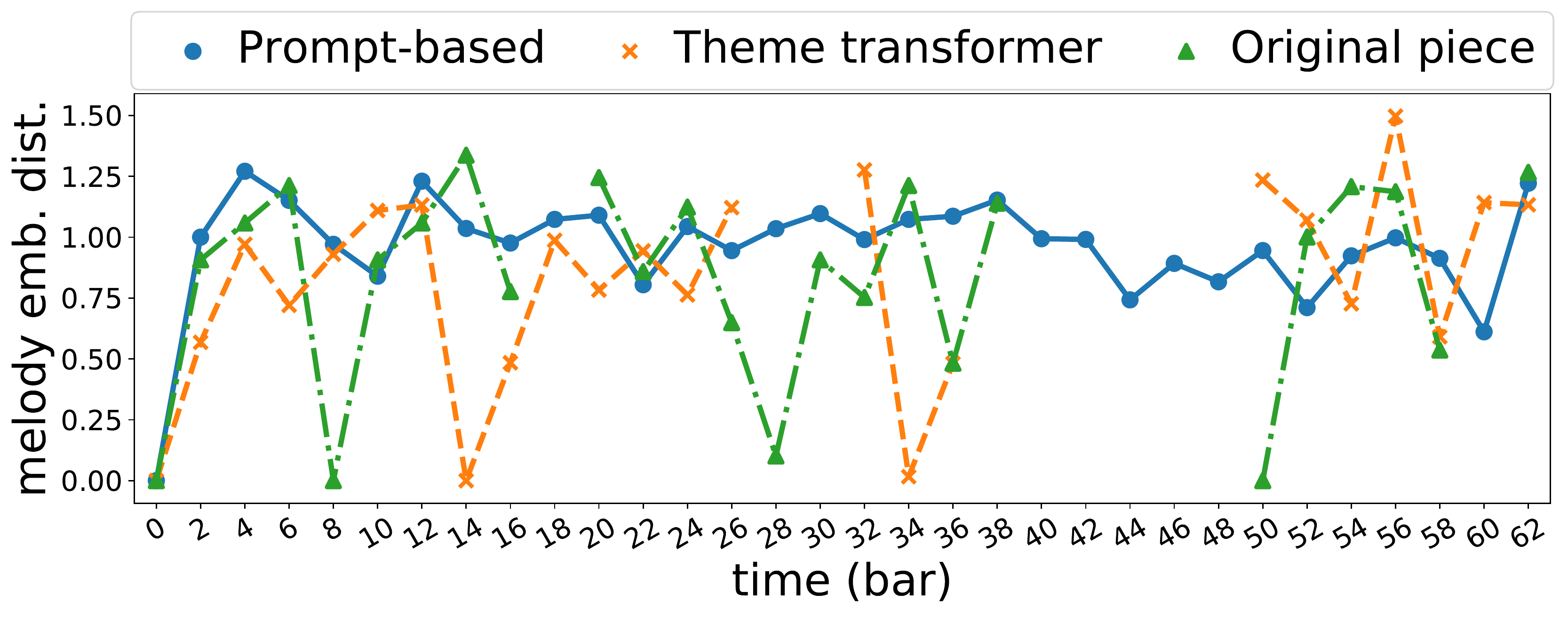} 
\\
\includegraphics[width=.45\textwidth]{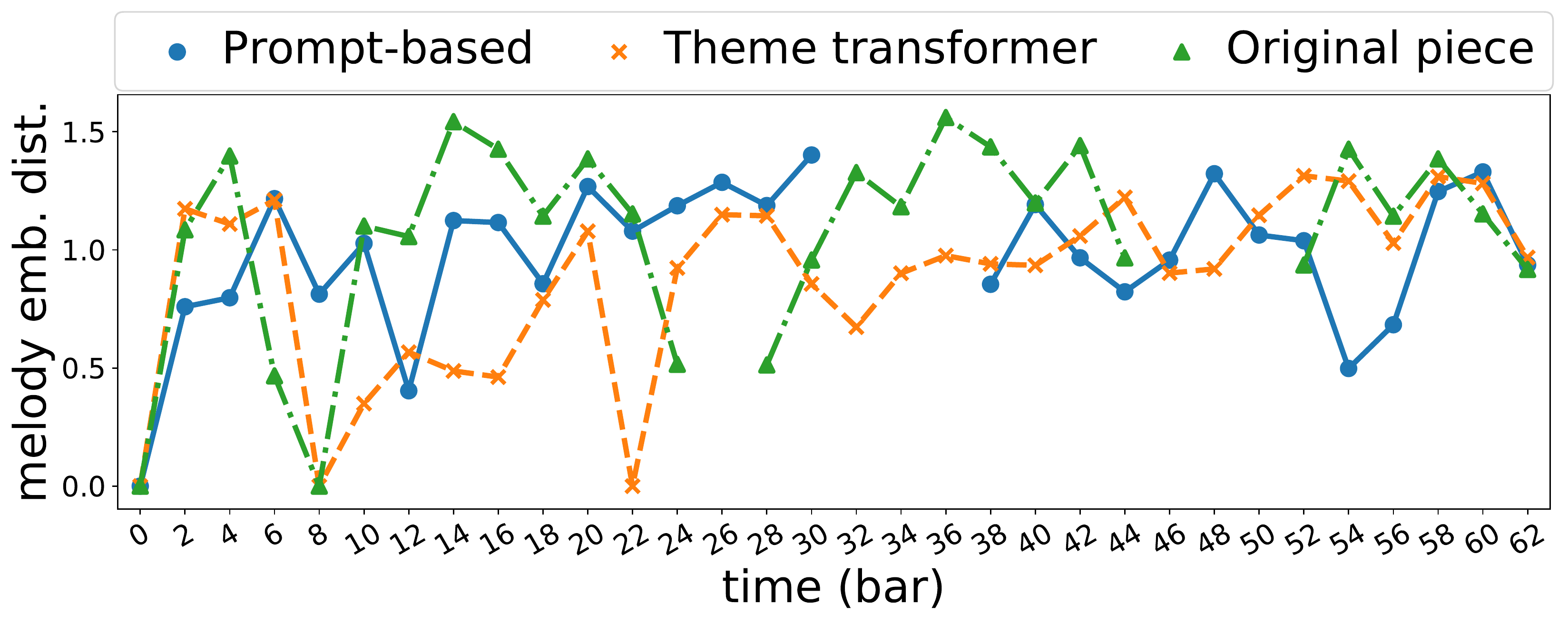}
&
\includegraphics[width=.45\textwidth]{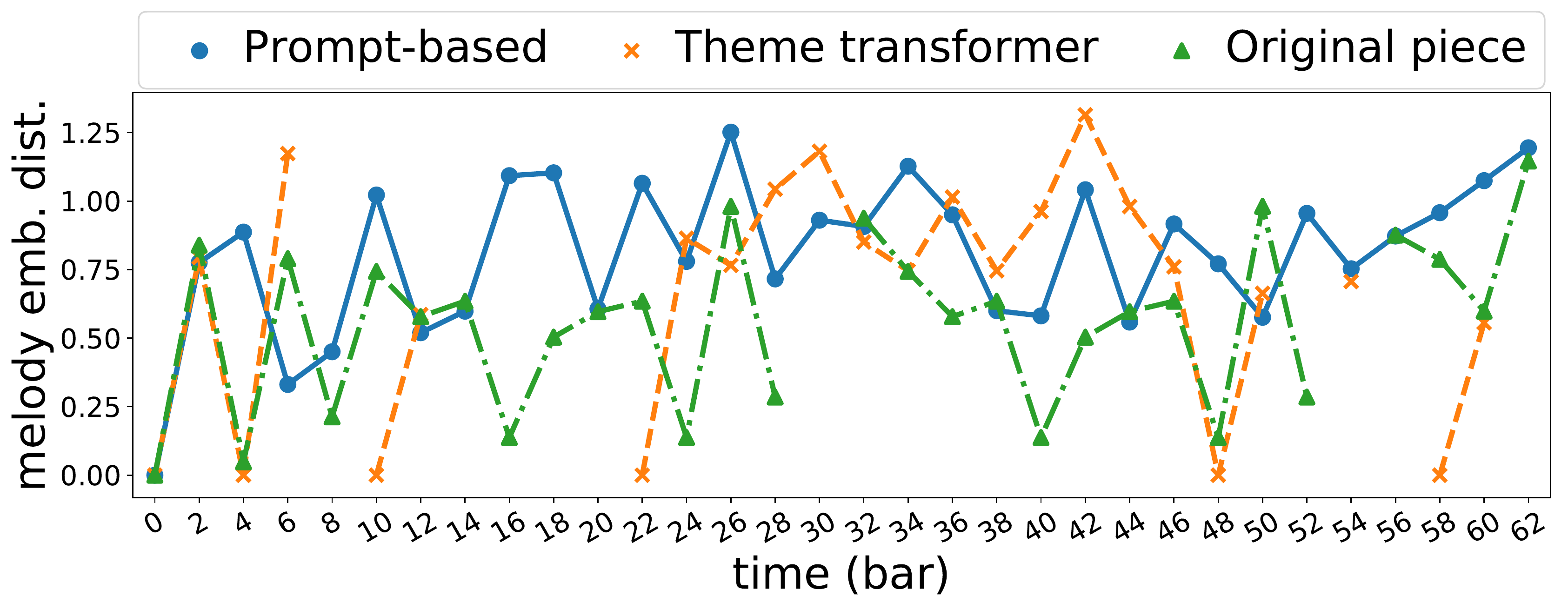}
\\
\includegraphics[width=.45\textwidth]{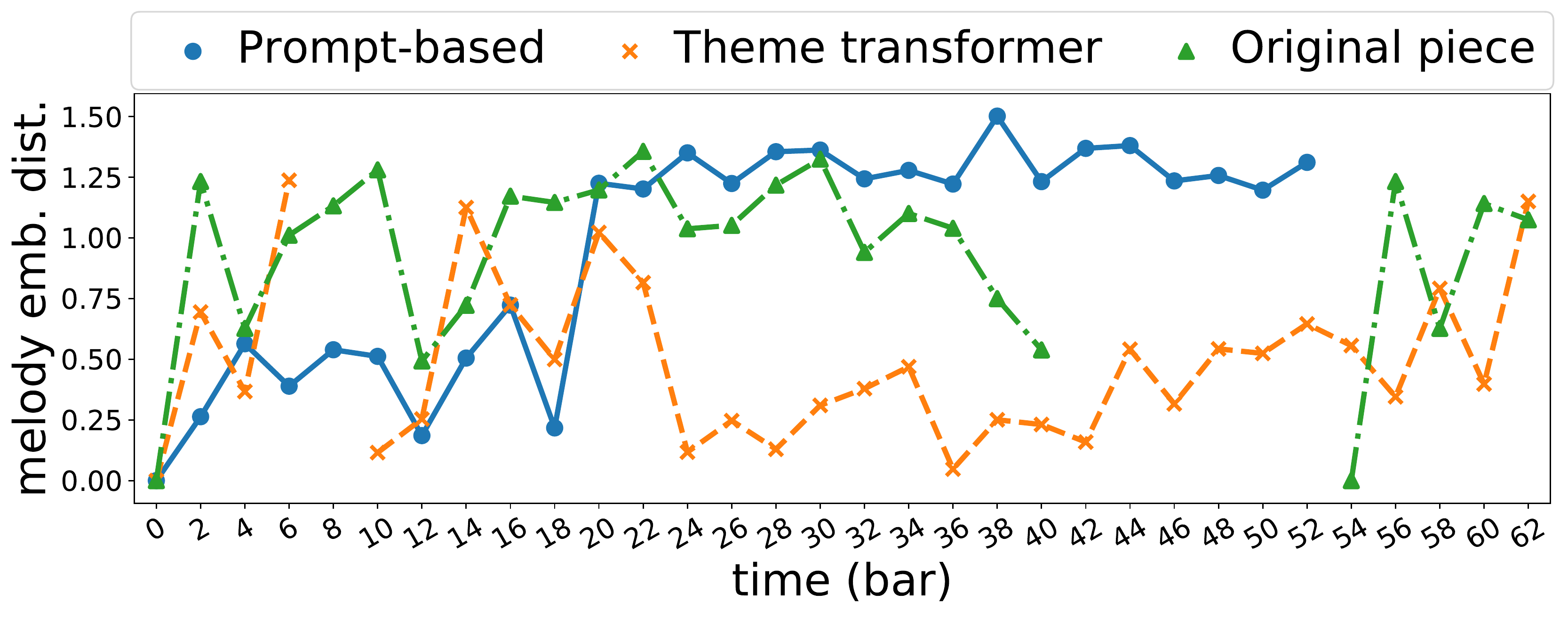}
&
\includegraphics[width=.45\textwidth]{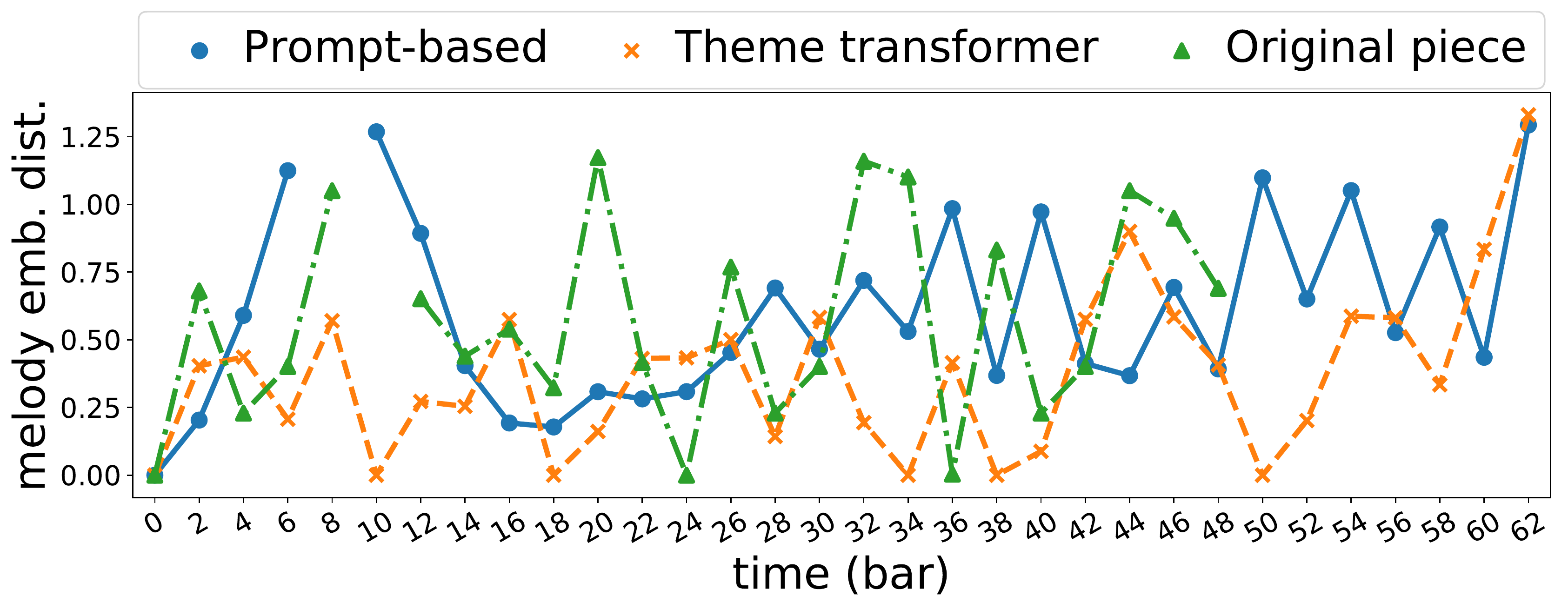} 
\\
\includegraphics[width=.45\textwidth]{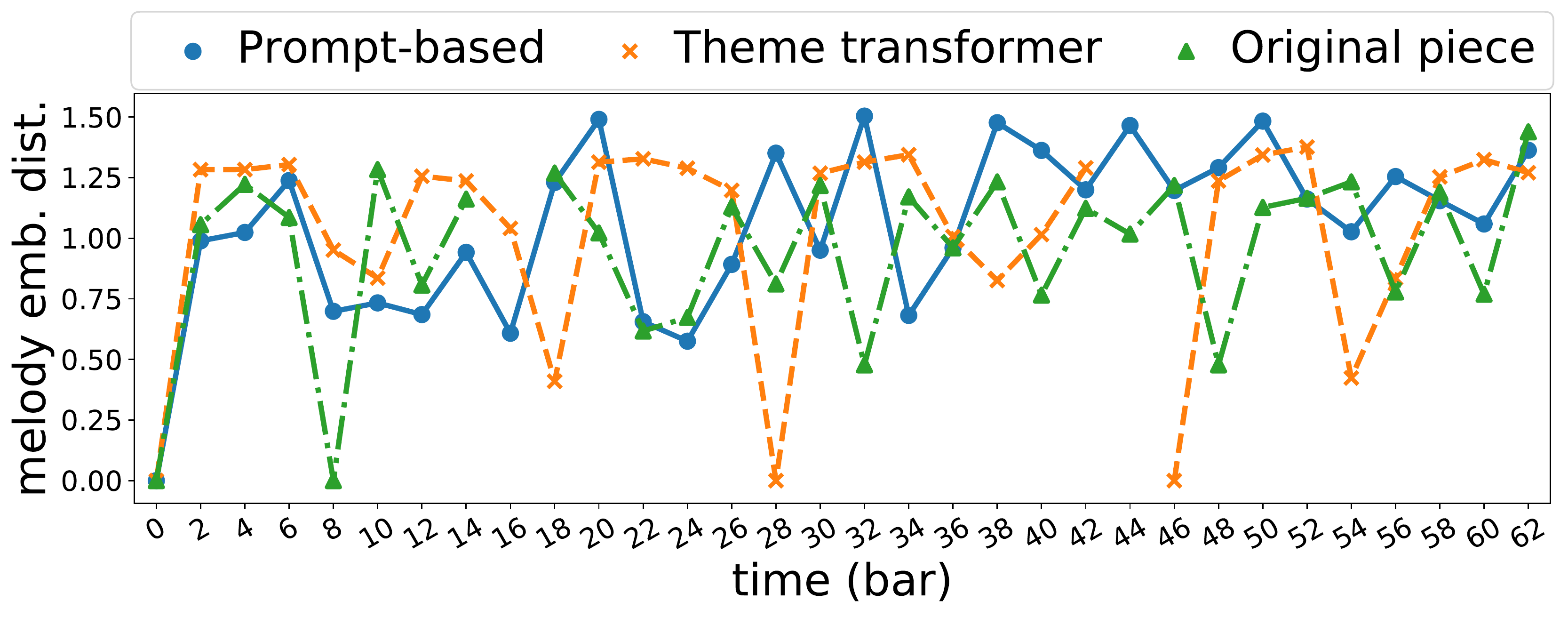}
&
\includegraphics[width=.45\textwidth]{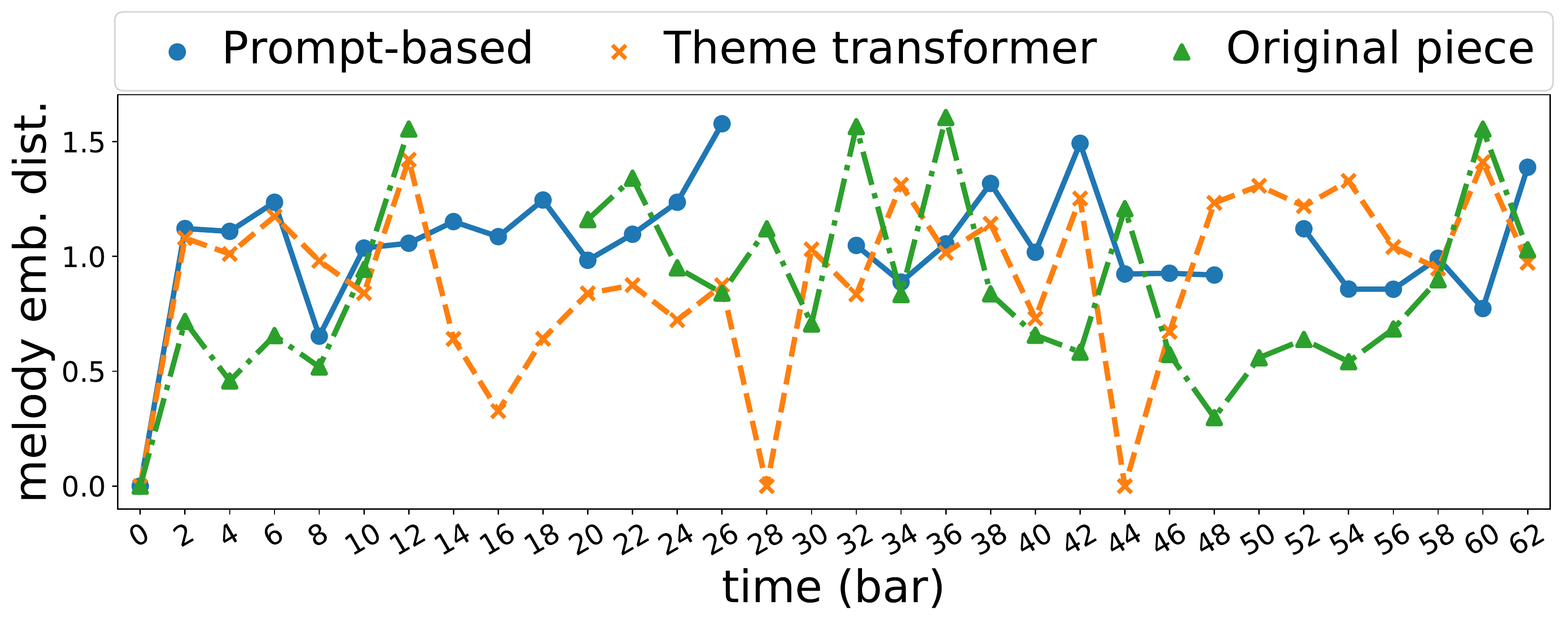}
\\
\includegraphics[width=.45\textwidth]{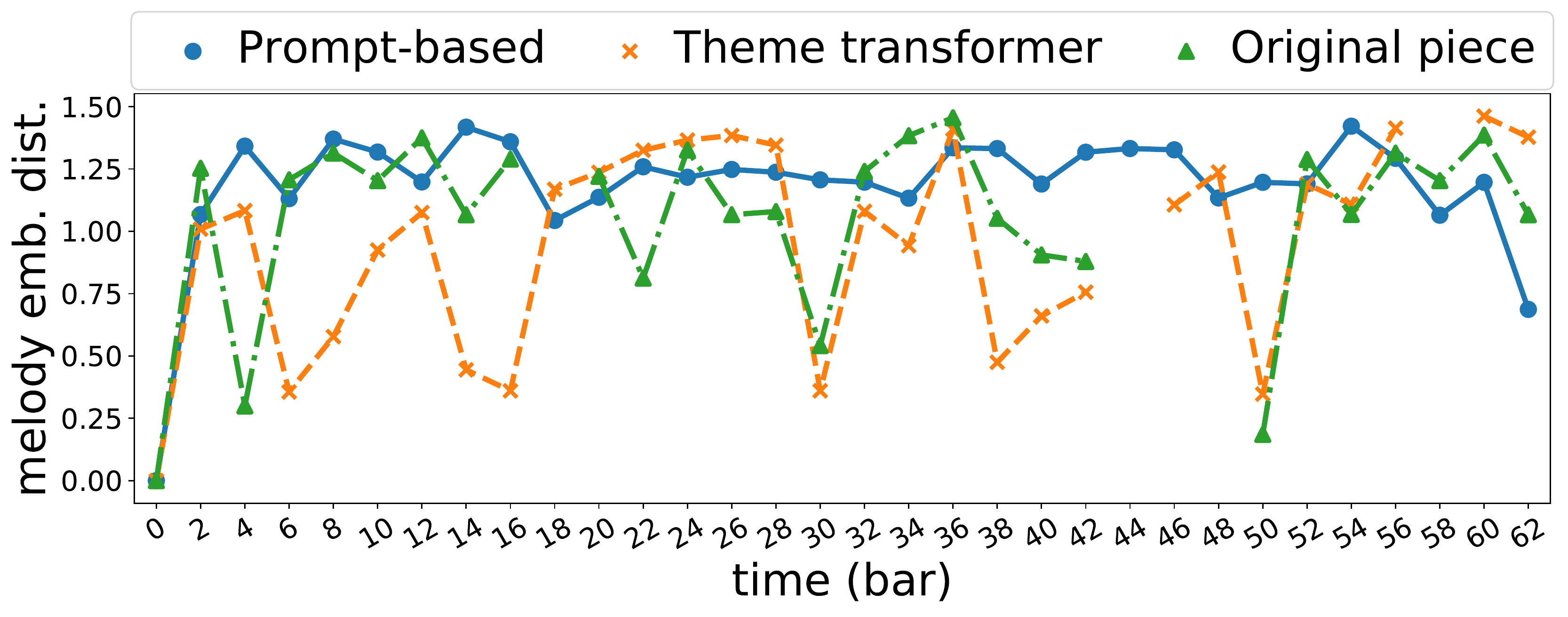} 
&
\includegraphics[width=.45\textwidth]{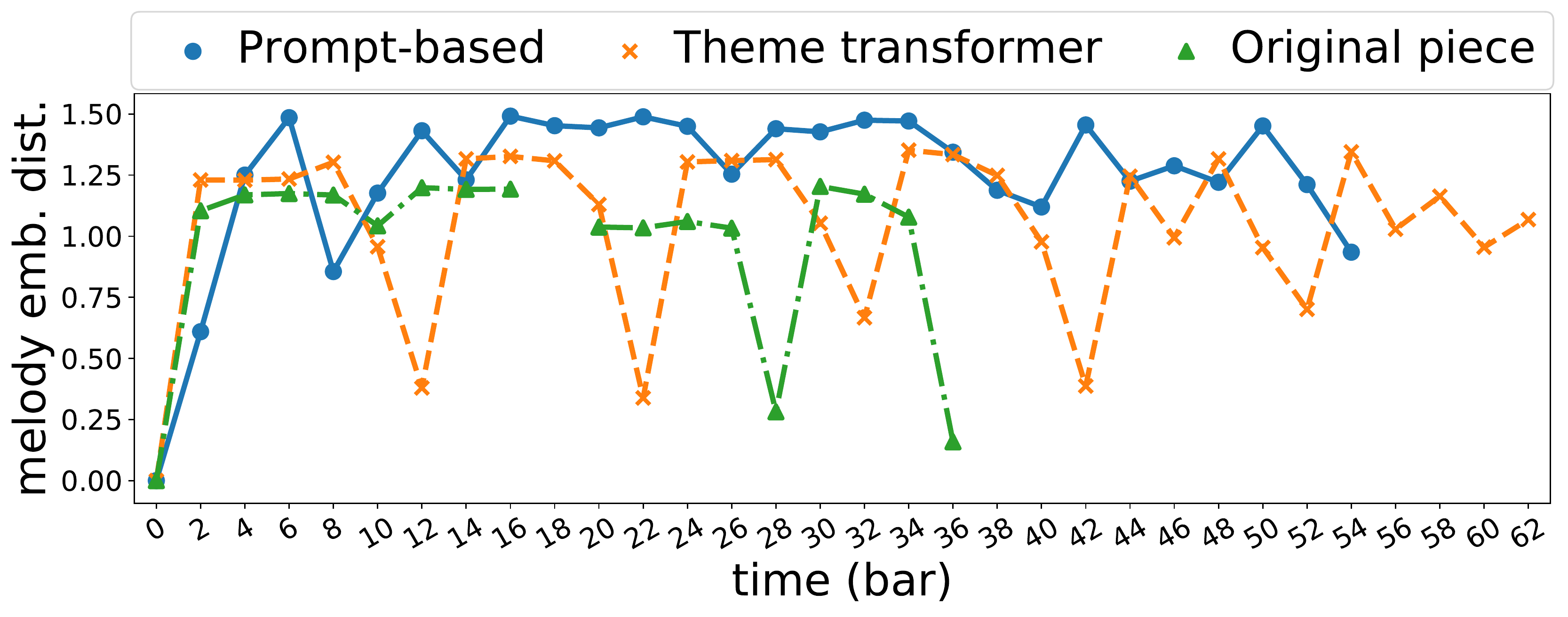}
\\
\end{tabular}
\caption{
What we draw in Figure \ref{fig:the-influence-of-cond-over-time} in the main body of the paper is actually about the \emph{melody inconsistency} (cf.  ``Objective Evaluation'' section) of an original or generated piece, i.e., $D(S_1,S_i)$, for $i\in[1,32]$, where $S_i$ denotes the $i$-th two-bar fragment of the piece (e.g., $S_1$ corresponds to the first two bars and $S_{32}$ the 63-rd and 64-th bars).
$D(\cdot,\cdot)$ is defined in Eq.~(\ref{eq:dist}).
We draw such figures for all the 12 testing songs participated in the listening test here (from left to right and then top to bottom: `875.mid' `888.mid'
`890.mid' `893.mid' `894.mid' `896.mid' `899.mid' `900.mid' `901.mid' `904.mid' `908.mid' `909.mid').
Those valleys in the curves, especially those below $\epsilon=0.13$, indicate fragments that are similar to $S_1$ (which is highly related to the conditioning theme) in their melodic content.
For fragments with no melody notes, $D(S_1,S_i)$ would return NaN, leading to discontinuities in the curves. In the last song, the original piece stops earlier because it is originally not that long. }
\label{fig:more-fig1}
\end{figure*}
\begin{figure*}
\includegraphics[width=\textwidth]{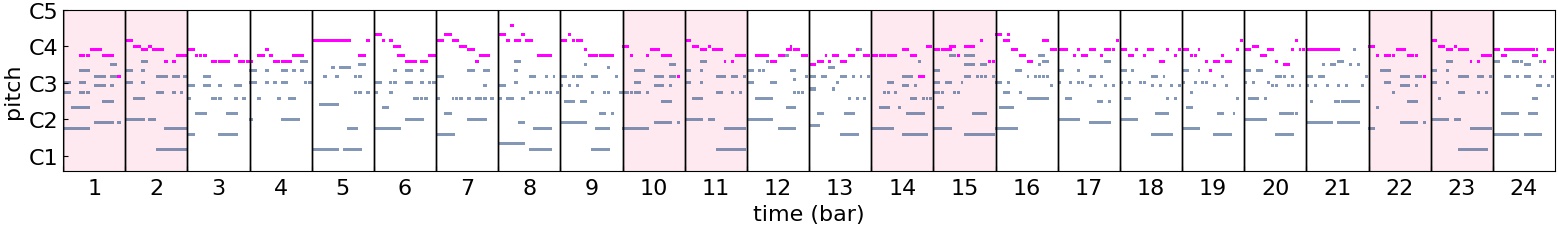} \\
\includegraphics[width=\textwidth]{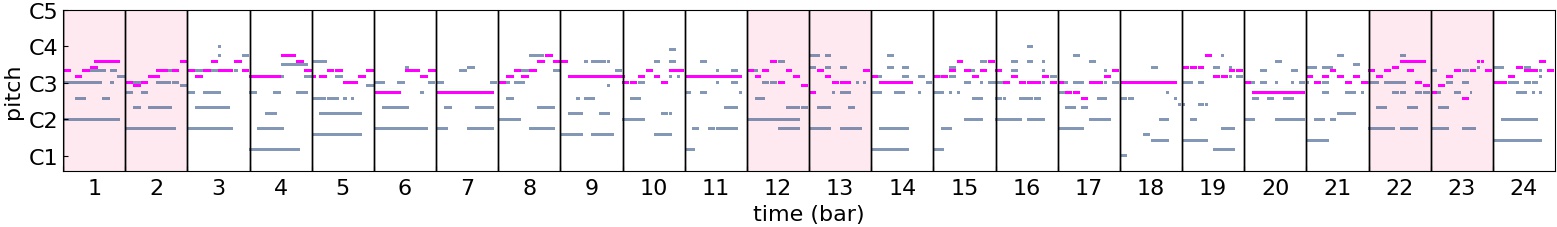} \\
\includegraphics[width=\textwidth]{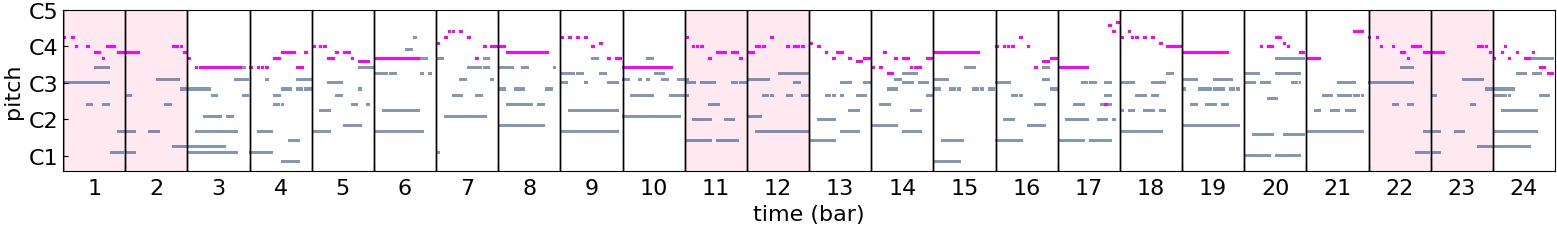} \\
\includegraphics[width=\textwidth]{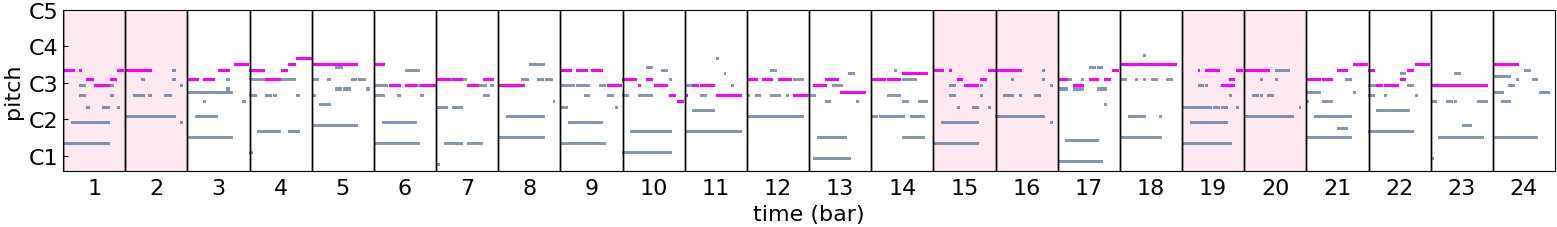} \\
\includegraphics[width=\textwidth]{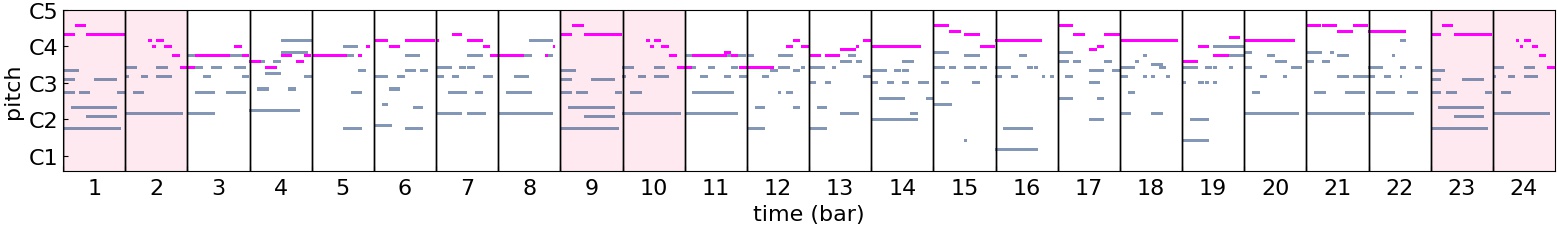} \\\includegraphics[width=\textwidth]{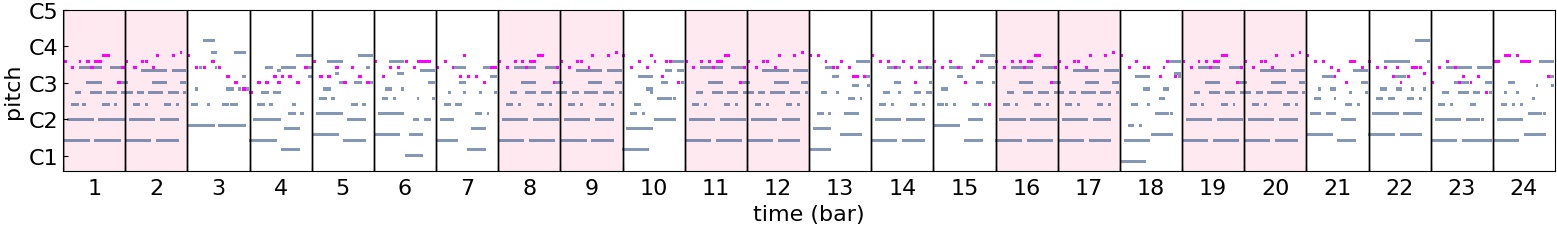} \\
\includegraphics[width=\textwidth]{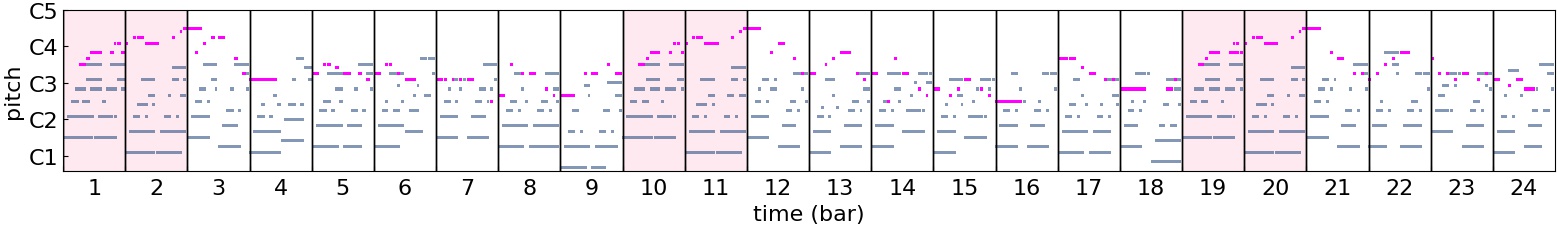} \\
\includegraphics[width=\textwidth]{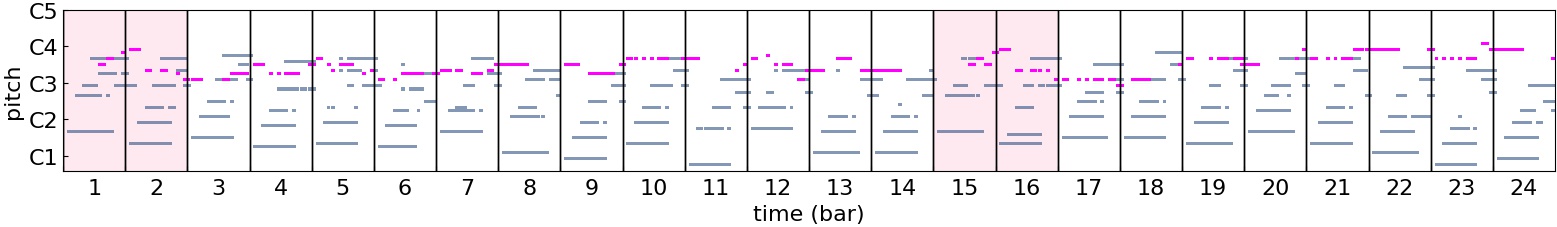}
\caption{The piano roll of the first 24 bars of a composition by Theme Transformer, conditioned on the theme of unseen testing songs
(from top to bottom : `875.mid' `888.mid'
`890.mid' `893.mid' `894.mid' `900.mid' `901.mid' `904.mid') from POP909. (Melody in magenta, accompaniment in grey, generated theme regions shaded in pink).}
\label{fig:more-fig-pianorolls}
\end{figure*}

Illustrations of examples of the piano music generated by the Theme Transformer can be found in Figures \ref{fig:more-fig1} and \ref{fig:more-fig-pianorolls}.

\subsection{Ablation Study}

We also report the result of an ablation study that probes the performance of variants of our model, to validate some of the design choices and to offer insights.


\subsubsection{On sequence length}

As discussed in our comments on the objective result on \emph{theme gap} in Section \ref{sec:obj:main}, 
the short sequence length  $N=512$ might have negatively affected the performance of our model in theme timing and structureness. 
Hence, we implement a variant of the Theme Transformer with $N=1024$. 
The second row of Table~\ref{tab:ablation} shows that, increasing the sequence length does improve the model's performance in \emph{melody inconsistency}, \emph{theme inconsistency} and \emph{theme uncontrollability}.
The mean and standard deviation in \emph{theme gap} also become much closer to those of the original pieces.
This suggests the potential to further improve the model 
by adopting a linear-complexity Transformer architecture \cite{katharopoulos2020transformers, choromanski2021rethinking} and using the whole song without slicing to train the Transformer, like we have attempted for the case of unconditional generation \cite{hsiao21aaai}. 

\subsubsection{On segment embedding vs. theme-aligned positional encoding}
In Theme Transformer, we use separate $\mathrm{PE}$s and no $\mathrm{SE}$, assuming that this works better. Here, we evaluate a variant of the Theme Transformer that uses the $\mathrm{SE}$ instead. Accordingly, the only difference between this variant to the basic seq2seq model would be the use of  gated parallel attention.
The third row of Table~\ref{tab:ablation} shows that this degrades the performance in all the metrics, 
in particular theme uncontrollability.
Moreover, a cross reference to Table \ref{tab:objective_result} shows that this variant outperforms the basic seq2seq model in \emph{melody inconsistency} and \emph{theme inconsistency} but not \emph{theme uncontrollability}.
This suggests that the proposed gated parallel attention contributes to the first two metrics, while the idea of separate $\mathrm{PE}$s contributes to the latter metric.

\subsubsection{On the number of trainable parameters}

Because of the additional encoder part of the Theme Transformer, the total number of trainable parameters of the Theme Transformer is roughly twice than that of the  baseline decoder-only model in the evaluations reported in the main body of the paper, which may not be fair. Therefore, we implement a variant of the prompt-based baseline that has $L=12$ self-attention layers in the decoder, to make the number of trainable parameters comparable to that of the Theme Transformer.
As Table~\ref{tab:ablation} shows, increasing the number of layers does not improve the result of the prompt-based baseline, suggesting that the high melody inconsistency is perhaps attributed to the inherent demerits of the prompt-based approach.

\begin{table*}[t]
\centering
\small
\begin{tabular}{|c|l|cccccc|} 
\hline
\multicolumn{2}{|c|}{} & \textbf{C}~\scriptsize{ontrol}  & \textbf{R}~\scriptsize{epeat} & \textbf{T}~\scriptsize{iming}  & \textbf{V}~\scriptsize{ariation} & \textbf{S}~\scriptsize{tructure} & \textbf{Q}~\scriptsize{uality}  \\ 
\hline \hline

&\iantext{Baseline (prompt-based)} \cite{huang2018music,huang20remitransformer} &3.01$\pm$1.08&2.55$\pm$1.18&2.73$\pm$1.06&2.65$\pm$1.06&3.06$\pm$0.94&3.19$\pm$0.98\\ 
User group 1
&Seq2seq \cite{huang2018music} &2.52$\pm$1.10&2.12$\pm$1.08&2.27$\pm$1.08&2.41$\pm$1.18&3.10$\pm$0.99&3.23$\pm$0.92\\ 
\scriptsize{(33 subjects)}
&Theme Transformer (proposed) &{3.63$\pm$1.10}&{3.55$\pm$1.22}&{3.27$\pm$1.03}&{3.03$\pm$1.11}&{3.33$\pm$0.99}&{3.38$\pm$0.97}\\ 
\hline\hline
&\iantext{Baseline (prompt-based)} \cite{huang2018music,huang20remitransformer} &2.90$\pm$1.09&2.39$\pm$0.97&2.76$\pm$1.26&3.22$\pm$1.24&2.78$\pm$1.09&2.78$\pm$1.00 \\ 
User group 2 
&Theme Transformer (proposed) &{3.49$\pm$1.11}&{3.39$\pm$1.12}&{3.27$\pm$1.25}&{3.25$\pm$1.06}&{3.16$\pm$1.00}&{3.16$\pm$1.00}\\
\scriptsize{(17 subjects)} 
& Original pieces &3.61$\pm$1.17&3.37$\pm$1.14&3.53$\pm$1.11&3.29$\pm$1.11&3.39$\pm$0.97&3.41$\pm$1.11 \\ 
\hline
\end{tabular}
\caption{Mean opinion scores (MOS) in the following six aspects across two different user groups: (C) theme controllability, (R) theme repetition, (T) theme timing, (V) theme variation, (S) overall structureness, (Q) overall quality. The first user group is 
familiar with Mandarin pop music, while the second group does not. 
}
\label{tab:subjective_result}
\end{table*}

\section{Qualitative Evaluation}

\subsection{Listening Test: Setup}
We establish a website and recruit 
anonymous human subjects to listen to a thematic material first, and then rate the correspondingly generated pieces in the following aspects, on a 5-point Likert scale (from 1 to 5; the higher the better):
\begin{itemize}
    \item {Theme \underline{c}ontrollability (C)}: How strongly is the piece influenced by the given theme?
    \item {Theme \underline{r}epetition (R)}: How often does the theme occur?
    \item {Theme \underline{t}iming (T)}: How well does the theme emerges in the piece at the timing you expect?
    \item {Theme \underline{v}ariation (V)}: How nicely does the theme vary in each of its occurrence in the piece?
    \item {Overall \underline{s}tructureness (S)} of the music piece.
    \item {Overall \underline{q}uality (Q)} of the music piece.
\end{itemize}
We intend to have subjects evaluate both the machine generated pieces and original pieces from the test set. However, some subjects might be familiar with the tunes in the original pieces, rendering the comparison unfair. Therefore, we require our subjects to self-report whether they are familiar with Mandarin pop music, and only ask those who are unfamiliar to listen to the original pieces.
For subjects of the first group (``familiar''), we ask them to rate the music generated by the three implemented models based on the thematic material of three songs randomly picked from the test set. For the second group (``unfamiliar''), we replace the music generated by the seq2seq model by the original pieces  (i.e., the testing songs). 
We randomize the order of the pieces corresponding to the same song and do not let the subjects know
whether there are any original pieces. 
But, we do tell them that our goal is to build a theme-conditioned music generation AI. 


The subjects are recruited from the social circles of the authors, which are from different countries. We ask the subjects to indicate (yes or no) whether they are familiar with Mandarin pop music by the question: ``Do you listen to Mandarin pop songs?''
We also ask them to optionally self-report their musical performance and composing experience.
We have examined whether their musical experience is related to their ratings but found nothing significant.
Among the subjects, 27 self-report themselves as male, and 23 as female.

Due to the time takes to set up the online survey website, we only use 12 out of the 29 testing songs in the listening test,  all selected randomly. The 12 songs are distributed in four packs, 
so that a subject listens to one of such packs.
For listeners, we convert each original and generated token sequence into a MIDI file, and then render it into audio by FluidSynth (\url{https://pypi.org/project/midi2audio/}), without additional mixing.\footnote{Please note that, the ``original pieces'' used in the objective and subjective evaluations are actually not the original MIDI files provided by POP909, but instead MIDI files converted from our token sequences. This is for a fair comparison between the original and generated pieces:
both are quantized to 1/4 beats,  no pedals, and no BRIDGE notes. Accordingly, they are actually ``degraded'' version of the original pieces.}

\subsection{Listening Test: Result}
Table~\ref{tab:subjective_result} shows that, from the result of both user groups, the Theme Transformer outperforms the other two models across all metrics, especially in the first three theme-related ones.
Both the prompt-based baseline and the basic seq2seq model cannot harness the condition well.
Besides, the Theme Transformer obtains similar MOS as the original data for all the theme-related metrics, demonstrating its effectiveness.


Moreover, there is a gap between the Theme Transformer and the original pieces in the last two ``overall'' metrics, calling for future improvement again.  
Interestingly, comparing the result of the overall metrics of the two user groups, it appears that the subjects can more easily notice the difference between the prompt-based baseline and the Theme Transformer with the original pieces as a reference. 


\subsection{Comments on the Quality of the Generated Music}
\label{sec:sub:comments}

With carefully examination of the music generated by our Theme Transformer, we found that each occurrence of the conditioning theme (i.e., a generated theme region) depends largely on the \emph{melody} notes in its preceding non-theme region.
When the preceding non-theme region has no melody notes at all, the subsequent theme-region tends to be highly similar to the given condition.
Otherwise, the proposed model brings in some variations of the theme that are related to the preceding melody notes.
It appears that our model does 
learn the difference between melody and accompaniment with relation to theme variation.

Our subjective assessment on \emph{theme controllability}, \emph{theme repetition}, and \emph{theme variation} in Table \ref{tab:subjective_result} shows that the Theme Transformer has learned to exploit the theme repetitively in its generation.
However, we found that the model fails to properly \textbf{develop the theme} throughout a generated piece, according to our own listening and also the score on \emph{overall structureness} in Table \ref{tab:subjective_result}.
Ideally, a song should follow some reasonable development, creating tensions, expectations, and giving listeners a sense of direction in melody variation.
But, our model tends to generate random variation of the theme in each theme region. 
We conjecture that this can be attributed to the short sequence length $N=512$, which amounts to about 9 bars of music only in  POP909. As shown in Table \ref{tab:objective_result}, in the original pieces the average \emph{theme gap} is 12.24 bars, meaning that a training sequence hardly contains two occurrences of the theme. The model cannot learn to take the preceding theme region into account while generating the next theme region. Moreover, due to  sequence slicing, the model does not know the position of each theme region (e.g., in the beginning or the end) in a training piece.

\begin{table*}[t]
\centering
\small
\iantext{
\begin{tabular}{|l|cc|ccc|ccc|} 
\hline\
& $\epsilon$ & $t$& Pitch class & Melody & Grooving  & Theme incon- & Theme uncon- & Theme gap \\ 
& & &consistency$\uparrow$&inconsistency$\downarrow$&consistency$\uparrow$&sistency$\downarrow$&trollability$\downarrow$& (in \# bars)\\
\hline \hline
\multirow{3}{*}{Theme Transformer}&
0.13&
1.2&
{.61$\pm$.06}&
{.13$\pm$.24}& 
{.92$\pm$.07}&
{0.27$\pm$0.26}& 
{0.24$\pm$0.20}& 
{9.48$\pm$3.59}\\ 
&
0.25&
1.2&
{.63$\pm$.05}&
{.23$\pm$.20}& 
{.91$\pm$.08}&
{0.42$\pm$0.23}& 
{0.66$\pm$0.42}& 
{8.41$\pm$3.05}\\ 
&
0.13&
1.8&
{.62$\pm$.07}&å
{.19$\pm$.25}& 
{.92$\pm$.06}&
{0.40$\pm$0.28}& 
{0.38$\pm$0.26}& 
{9.43$\pm$3.56}\\ 
\hline
\multirow{2}{*}{Original pieces} 
& 
0.13&
--&
.65$\pm$.05& 
.09$\pm$.18&
.74$\pm$.10& 
0.05$\pm$0.05& 
0.04$\pm$0.04& 
12.24$\pm$11.32\\
& 
0.25&
--&
.65$\pm$.05& 
.09$\pm$.18&
.74$\pm$.10& 
0.31$\pm$0.27& 
0.57$\pm$0.45& 
9.91$\pm$9.29\\
\hline
\end{tabular}
}

\caption{\iantext{Ablation results for different $\epsilon$ used in DBSCAN and different temperature ($t$) used during sampling. }}
\label{tab:other_result}
\end{table*}

\section{\iantext{Discussion}}

\subsection{\iantext{On the Variance of the Generated Theme Regions}}
\label{sec:sub:commentsVar}
\iantext{
Table~\ref{tab:other_result} shows that we can further increase the variance of generated theme regions by increasing the threshold $\epsilon$ in the clustering algorithm DBSCAN and re-training the Theme Transformer.  
Increasing the value of  $\epsilon$ allows higher diversity among the segments grouped to the same cluster, and thereby better informs the model what a reasonable variation is at the training time.
We can see from the last two rows of Table~\ref{tab:other_result} that increasing $\epsilon$ from 0.13 to 0.25 indeed increases the inconsistency of the theme regions in the original pieces in the training data from 0.05$\pm$0.05 to 0.31$\pm$0.27. 
The second row of Table~\ref{tab:other_result} shows that, the theme inconsistency of the music generated by the re-trained Theme Transformer (with $\epsilon=0.25$) is indeed higher than the original Theme Transformer ($\epsilon=0.13$), but the obtained theme inconsistency (0.42$\pm$0.23) remains close to that of the original data (0.31$\pm$0.27). Our informal listening shows that setting $\epsilon$ larger contributes to higher theme variance without much hurting musical quality.

In contrast, simply increasing the temperature $t$ of the Transformer decoder during the inference stage (e.g., from 1.2 to 1.8, as shown in Table~\ref{tab:other_result}) cannot effectively increase the variance of the generated theme regions. 
Doing so mainly introduces randomness in the inference stage,  
generating music that might contain unpleasant note combinations that degrade the overall musical quality, according to our informal listening.

}

\subsection{\iantext{On the Quality of Theme Retrieval}}
\label{sec:sub:commentsThmRetr}
\iantext{

To have an idea of the performance and limit of the proposed ``contrastive learning$+$clustering'' based theme retrieval method (denoted as \textit{CL} hereafter), we invite three annotators with some musical training to annotate the theme and its occurrences of six random songs from the test set of POP909.\footnote{\iantext{We are aware of two other relevant dataset, the MTD \cite{ZalkowBAM20MTDTISMIR} and JKU Pattern Dataset \cite{benamar17,janssen13CMMRa,LemstromSIA,DavidMRECURSIA}. However, the former has no labeling of the occurrences of the theme in the entire music piece, and the latter is about musical patterns, which can be of arbitrary length (i.e., can be too short or too long) and do not necessarily correspond to musical themes. Moreover, both of them consist only of classical music. As a result, we need to create our own labeled data to quantify the performance of theme retrieval.}}
We have three annotators to account for the possible subjectivity of the theme labeling.
We can then calculate the F1-score, which is the harmonic mean of the recall and precision rates, of our theme retrieval method, using each time the labels from one of the annotators as the ``ground truth.''
We explain to the annotators the meaning of themes and 
instructed them that 
1) the labeling is by nature subjective so they are feel to use their own judgement; 
2) there is no need to align either the beginning or the end of a theme region to the bar lines;
3) a theme region should roughly correspond to a music phrase and its length might be about 2 to 4 bars.
The annotators use their own MIDI editor to annotate the beat indices that they consider belonging to the theme regions.


We evaluate not only CL but also the following methods:
\begin{itemize}
    \item \textit{CL w/o Note Dur.}: the ablated version of CL with only the ``Pitch shift on scale'' augmentation.
    \item \textit{CL w/o Pitch Shift}: the ablated version of CL without the ``Pitch shift on scale'' augmentation.
    \item \textit{Correlative-matrix} (CM)\cite{HsuNonTrivialPattern}, a string-based musical pattern mining method that
    represents melody pitches in morphetic symbols and searches for closed patterns in the resulting melody string. It considers the longest sub-string that has less than 30 notes and that repeats at least twice in a song as the theme.
    \item \textit{COSIATEC}\cite{LemstromSIA}, a geometric-based musical pattern mining method that views the melody notes as dots in a 2-dimensional space of \{onset, pitch\} and then extracts the 
    ``translational equivalence class'' (TEC) groups from the melody. 
    We choose the TEC group with the highest compactness ratio in a song as the theme. Moreover, we remove the patterns with overlapping time span, since the theme regions should not collide with each other.
\end{itemize}
We provide the MIDI files and pianoroll-like graphical visualization of the theme regions labeled by the three annotators and predicted by all the implemented methods at  \url{https://atosystem.github.io/ThemeTransformer}. 

The average F1 scores for the implemented methods across the six songs and three annotators turn out to be 0.378 (CL), 0.220 (CL w/o Note Dur.), 0.336 (CL w/o Pitch Shift), 0.345 (CM), and 0.297 (COSIATEC), respectively. We see that CL indeed reaches the highest average F1 score, but the two-sided paired t-test shows that the performance difference between CL and any other methods is not significant from a statistical point of view, except for the difference between CL and `CL w/o Not Dur.' ($p$-value$<$0.05).

We further perform a qualitative analysis of the theme retrieval results, 
finding that the methods CL, CM and COSIATEC actually have their strengths and weaknesses and it is hard to say which method performs the best. 
A strength of CL is that we can easily alter the theme variation of a song by tuning the $\epsilon$ parameter of DBSCAN (see Section \ref{sec:sub:commentsVar}), which is not available for CM and COSIATEC.
However, we found qualitatively that the performance of CL is largely limited by the na\"ive melody segmentation method we adopted---as described in Section \ref{sec:theme-retrieval:segmentation}, we simply sliced a song into two-bar fragments without overlaps at the bar lines, and used those two-bar fragments as candidates of the theme regions. 
As a result, for some songs the theme regions might be incomplete chord progression or minor melody, missing the essential part.
Both CM and COSIATEC do not have this two-bar segmentation assumption and can sometimes retrieve segments with musically better sense than CL.


We note that, however, the general idea of combining contrastive learning and clustering for theme retrieval can work with arbitrary melody segmentation method.
While we adopt a na\"ive melody segmentation method in our current implementation, we should investigate replacing it with a better melody segmentation method in the future, ideally a method that can accurately segment a music piece into musically meaningful phrases. This holds the promise of further improving the performance of Theme Transformer.


}

\section{Conclusion}

In this paper,
we used contrastive learning with musically-inspired data augmentation and DBSCAN clustering to extract thematic materials occuring throughout a piece.
We then proposed Theme Transformer, an encoder-decoder model that takes advantage of gated parallel attention and theme-aligned positional encoding to refer to the thematic material given as a condition during music generation.
Conducted objective and subjective evaluations demonstrate that, compared to the decoder-only Music Transformer and the simple Seq2seq Transformer,  Theme Transformer does better 
in repeating the thematic materials with perceptible variations, 
and 
enhances the quality of the generated music.




This work can be extended in many ways.
First, we only considered composing with just one given theme.
However, 
\iantext{as mentioned in Section \ref{sec:sub:commentsThmRetr},}
a music piece (e.g., a sonata, or a pop song in verse-chorus form) can often contain multiple themes, which may be similar or in stark contrast to each other.
We may build a model that takes multiple themes as conditions, with mechanisms to alternate its attention between different themes to compose a well-formed piece.
Second, as seen in 
Figure~\ref{fig:the-influence-of-cond-over-time}, the similarity measure obtained from contrastive learning could be a great tool to characterize how the music ``departs from'' and ``returns to'' the theme in a piece.
This opens the door to developing methods that take such ``similarity curves'' into account during generation, thereby improving the development of the themes in the music, as well as offering users a medium to interact with music generation models.
Finally, 
we are interested in letting the model take a user-created theme as condition or generate the theme on its own from scratch, rather than extracting from existing pieces.

\bibliographystyle{IEEEtran}
\bibliography{main}
\vfill

\end{document}